\documentclass[sn-basic]{sn-jnl}

\newcommand{\edit}[1]{#1}

\jyear{2023}

\begin{document}

\title[Surface Flux Transport \edit{on the Sun}]{Surface Flux Transport \edit{on the Sun}}

\author*[1]{\fnm{Anthony R.} \sur{Yeates}}\email{anthony.yeates@durham.ac.uk}
\author[2]{\fnm{Mark C.~M.} \sur{Cheung}}\email{mark.cheung@csiro.au}
\author[3]{\fnm{Jie} \sur{Jiang}}\email{jiejiang@buaa.edu.cn}
\author[4]{\fnm{Kristof} \sur{Petrovay}}\email{k.petrovay@astro.elte.hu}
\author[5]{\fnm{Yi-Ming} \sur{Wang}}\email{yi.wang@nrl.navy.mil}

\affil*[1]{\orgdiv{Department of Mathematical Sciences}, \orgname{Durham University}, \orgaddress{\city{Durham}, \country{UK}}}
\affil[2]{\orgname{CSIRO, Space \& Astronomy}, \orgaddress{\city{Marsfield}, \state{NSW}, \country{Australia}}}
\affil[3]{\orgdiv{School of Space and Environment}, \orgname{Beihang University}, \orgaddress{\city{Beijing}, \country{People's Republic of China}}}
\affil[4]{\orgdiv{Department of Astronomy}, \orgname{E\"otv\"os Lor\'and University}, \orgaddress{\city{Budapest}, \country{Hungary}}}
\affil[5]{\orgdiv{Space Science Division}, \orgname{Naval Research Laboratory}, \orgaddress{\city{Washington}, \state{DC}, \country{USA}}}

\abstract{We review the surface flux transport model for the evolution of magnetic flux patterns on the Sun's surface. Our underlying motivation is to understand the model's prediction of the polar field (or axial dipole) strength at the end of the solar cycle. The main focus is on the ``classical'' model: namely, steady axisymmetric profiles for differential rotation and meridional flow, and uniform supergranular diffusion. Nevertheless, the review concentrates on recent advances, notably in understanding the roles of transport parameters and -- in particular -- the source term. We also discuss the physical justification for the surface flux transport model, along with efforts to incorporate radial diffusion, and conclude by summarizing the main directions where researchers have moved beyond the classical model.
}


\maketitle

\section{Introduction}\label{sec:intro}

The surface flux transport (hereafter SFT) model is based on an elegant and simple idea, originally formulated by \cite{Leighton1964}: radial magnetic flux on the solar surface behaves like a passive scalar field. In other words, flux is carried around by horizontal plasma flows but with no back reaction on these flows.

\begin{figure}
    \centering
    \includegraphics[width=\textwidth]{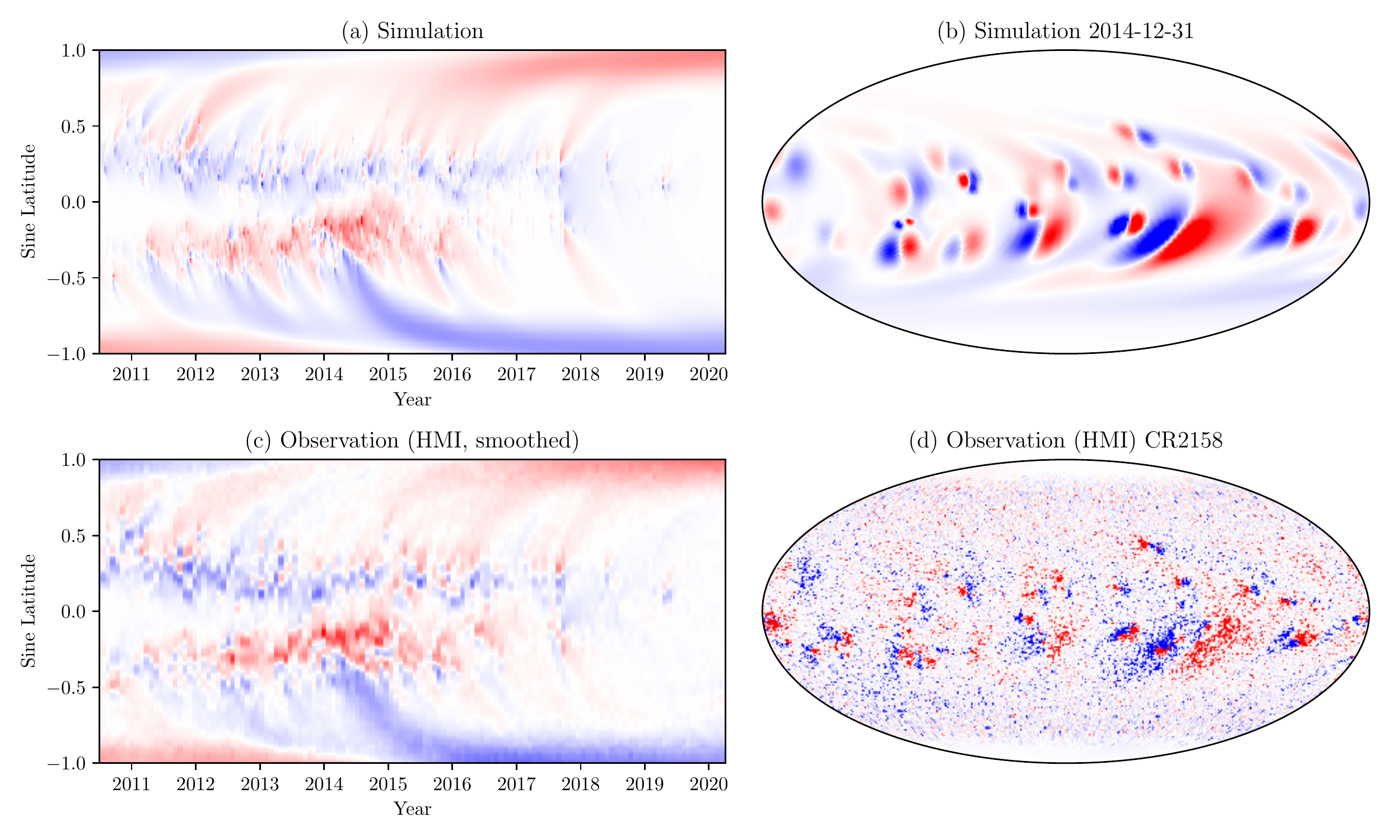}
    \caption{An SFT model for Solar Cycle 24 with emerging regions derived from SDO/HMI SHARPs data \cite[following the method of][]{Yeates2020}. Panel (a) shows the longitude-averaged field $\langle B_r\rangle$ in the simulation, and (b) shows a snapshot of the two-dimensional field $B_r$ on 31 December 2014. For comparison, (c) shows a magnetic butterfly diagram (or super-synoptic map) constructed from SDO/HMI pole-corrected synoptic maps \citep{Sun2018}, smoothed to a comparable resolution to the simulation. The individual, unsmoothed synoptic map for Carrington rotation CR2158 is shown in (d). Red/blue denote positive/negative values, capped at $\pm10\,\mathrm{G}$ in (a,c) and $\pm50\,\mathrm{G}$ in (b,d).}
    \label{fig:paper_opt_bfly}
\end{figure}

Despite its simplicity, the SFT model has proven remarkably successful at replicating the magnetic flux patterns on the real solar surface (photosphere). Figure \ref{fig:paper_opt_bfly} shows an SFT simulation for Solar Cycle 24, where new active regions have been inserted based on magnetograph observations. With appropriate parameters, the time-latitude ``magnetic butterfly diagram'' in the SFT model (Figure \ref{fig:paper_opt_bfly}a) is a good match for the observed time-latitude plot (Figure \ref{fig:paper_opt_bfly}c) at all latitudes. In general, the success of the SFT model has led to important applications both as (i) an inner boundary condition for extrapolations of the magnetic field in the solar atmosphere, and (ii) an outer boundary constraint on models for the solar interior dynamo.

In this review, our focus is on understanding the model itself: both its key ingredients and fundamental behaviour when applied in the solar regime. Details about applications, particularly to the solar atmosphere, may be found in previous review articles \citep{Sheeley2005, Mackay2012, Wang2017}. In the solar dynamo context, the SFT model has been used to constrain theories and models of the magnetic field in the solar interior \citep[e.g.,][]{Cameron2012b, Cameron2015, Jiang2014b,  Lemerle2017, Whitbread2019, Hazra2021}. But it is also a valuable practical tool for solar cycle prediction, enabling predictions to be made of the polar field at the end of the current solar cycle, and hence -- through well-established correlations -- the amplitude of the following solar activity cycle \citep[e.g.,][]{Cameron2016,Iijima2017,Jiang2018, Upton2018, Bhowmik2018, Jiang2022}. Understanding the origin and limitations of such polar field predictions requires an understanding of the SFT model itself, which is what we seek to provide here.

The review is organised as follows. In Section \ref{sec:fundamentals}, we present the basic equations of the ``classical'' SFT model.
Section \ref{sec:flows} discusses the imposed flows in the model, including the importance of including meridional flow and recent work on constraining the flow parameters.
Section \ref{sec:source} discusses the source term representing new flux emergence, which is fundamental to the flux patterns that the model predicts.
Section \ref{sec:justification} examines the important question of whether the SFT model -- usually seen as purely phenomenological -- can be derived from physical principles.
We conclude in Section \ref{sec:beyond} with an overview of model features beyond our ``classical'' version.

\section{Fundamentals of the Classical Model}\label{sec:fundamentals}

Denoting the radial magnetic field distribution by $B_r(\theta,\phi,t)$, the equation for a passive scalar field is
\begin{equation}
    \frac{\partial B_r}{\partial t} + \nabla_h\cdot\big({\bf u}_h B_r\big) = \eta\nabla_h^2B_r + S,
    \label{eqn:sft}
\end{equation}
where ${\bf u}_h$ is the imposed advection velocity, and $\eta$ is the diffusivity. In the classical model, $B_r$ represents the large-scale mean field\edit{; the model does not resolve the smaller-scale motions of supergranular convection, but rather models these with the turbulent diffusivity $\eta$. This was introduced by \cite{Leighton1964} to parameterise the ``random walk'' of individual magnetic flux elements due to the changing pattern of supergranular flows.} For SFT it is necessary to include also a prescribed source term $S(\theta,\phi,t)$ that describes the emergence of new magnetic flux, typically in the form of active regions. In a more complete physical model, $S$ would arise self-consistently through Faraday's induction equation (to be discussed in Sections \ref{sec:justification} and \ref{sec:beyond}), but in the classical SFT model it is a prescribed model input. Throughout we will use subscript $h$ to denote the ``horizontal'' components of a vector, meaning those tangential to the solar surface.

In the classical SFT model, the diffusivity $\eta$ is uniform and constant, and most authors assume a steady, axisymmetric imposed velocity of the form
\begin{equation}
    {\bf u}_h(\theta) = R_\odot\sin\theta\,\Omega(\theta){\bf e}_\phi + u_\theta(\theta){\bf e}_\theta.
    \label{eqn:uaxi}
\end{equation}
Thus $\Omega(\theta)$ represents the angular velocity of solar differential rotation, and $u_\theta(\theta)$ represents the meridional circulation. The choice of these flows is important and will be discussed further in Section \ref{sec:flows}. Relaxing the classical assumptions is considered in Section \ref{sec:beyond} (except for the addition of an exponential decay term which is discussed in Section \ref{sec:justification}).

\subsection{Dimensionless Form}

Ignoring $S$, we can consider non-dimensionalization of equation \eqref{eqn:sft} by defining dimensionless variables ${\bf u}'_h = {\bf u}_h/U_0$, $\nabla_h'=R_\odot\nabla_h$ and $t' = tU_0/R_\odot$, where $U_0$ is a typical flow speed. Then \eqref{eqn:sft} becomes
\begin{equation}
    \frac{\partial B_r}{\partial t'} + \nabla'_h\cdot\big({\bf u}_h' B_r\big) = \frac{1}{\mathrm{Rm}}{\nabla_h'}^2B_r,
    \label{eqn:sftnd}
\end{equation}
suggesting that the behaviour (in the absence of new emergence) is controlled by the dimensionless magnetic Reynolds number
\begin{equation}
    {\rm \mathrm{Rm}} = \frac{R_\odot U_0}{\eta}.
    \label{eqn:Lambda}
\end{equation}
In effect, it is only the relative speed of advective to diffusive transport that matters.

\subsection{Explicit Form}

Writing out \eqref{eqn:sft} explicitly in spherical coordinates, and assuming \eqref{eqn:uaxi}, gives the standard SFT equation
\begin{align}
    &\frac{\partial B_r}{\partial t} + \frac{1}{R_\odot\sin\theta}\frac{\partial}{\partial\theta}\Big(\sin\theta\,u_\theta B_r\Big) + \Omega(\theta)\frac{\partial B_r}{\partial\phi} = \nonumber\\
    &\qquad\qquad\qquad \frac{\eta}{R_\odot^2\sin\theta}\frac{\partial}{\partial\theta}\left(\sin\theta\frac{\partial B_r}{\partial\theta}\right) + \frac{\eta}{R_\odot^2\sin^2\theta}\frac{\partial^2 B_r}{\partial\phi^2} + S.
    \label{eqn:sft2d}
\end{align}
In some applications it suffices to consider the longitude-averaged field, 
\begin{equation}
    \langle B_r\rangle(\theta, t) = \frac{1}{2\pi}\int_0^{2\pi} B_r(\theta,\phi,t)\,\mathrm{d}\phi.
\end{equation}
Integrating \eqref{eqn:sft2d}, we find that $\langle B_r\rangle$ obeys the one-dimensional equation
\begin{align}
    \frac{\partial\langle B_r\rangle}{\partial t} + \frac{1}{R_\odot\sin\theta}\frac{\partial}{\partial\theta}\Big(\sin\theta\,u_\theta \langle B_r\rangle\Big) = \frac{\eta}{R_\odot^2\edit{\sin\theta}}\frac{\partial}{\partial\theta}\left(\sin\theta\frac{\partial\langle B_r\rangle}{\partial \theta}\right) + \langle S\rangle,
    \label{eqn:sft1d}
\end{align}
showing in particular that differential rotation has no effect on the evolution of $\langle B_r\rangle$ \citep{Leighton1964}. On the other hand, the differential rotation -- being the fastest flow -- plays an important role in determining the two-dimensional flux patterns seen on the solar surface. By increasing the length of the polarity inversion lines in and between active regions, it also speeds up the diffusive cancellation of non-axisymmetric components of $B_r$ \citep{Sheeley1986}.

\subsection{Implementation}

Although some analytical analysis is possible \citep[see][and also Section \ref{sec:source} below]{Sheeley1986, DeVore1987}, for most applications it is usual to solve \eqref{eqn:sft2d} or \eqref{eqn:sft1d} with numerical methods. This dates right back to the original paper of \cite{Leighton1964}. The most natural numerical approach would be a spectral method based on spherical harmonics, as implemented for example by \cite{Mackay2002} or \cite{Baumann2004} \citep[see][for more details]{Baumann2005}.
However, care is needed in treating the source term $S$, since newly-emerging active regions are typically highly localized in space and usually require filtering in spectral space to avoid the Gibb's phenomenon (``ringing''). A more straightforward approach is to use a simple explicit finite-volume method, \edit{provided that care is taken in both the discretization and the source term to} conserve magnetic flux (i.e., preserve $\int_0^{2\pi}\int_0^\pi B_r(\theta,\phi,t)\sin\theta\,\mathrm{d}\theta\mathrm{d}\phi = 0$).  The resulting time-step restriction is typically not a severe problem on modern machines, given the two-dimensional nature and modest resolutions typically used (for example, a  $360\times 180$ mesh). Much higher resolutions would not be consistent with the mean-field assumption of the classical model (alternatives are discussed in Section \ref{sec:beyond}).

\section{Flows}\label{sec:flows}

\subsection{Differential Rotation}

\begin{figure}
    \centering
    \includegraphics[width=\textwidth]{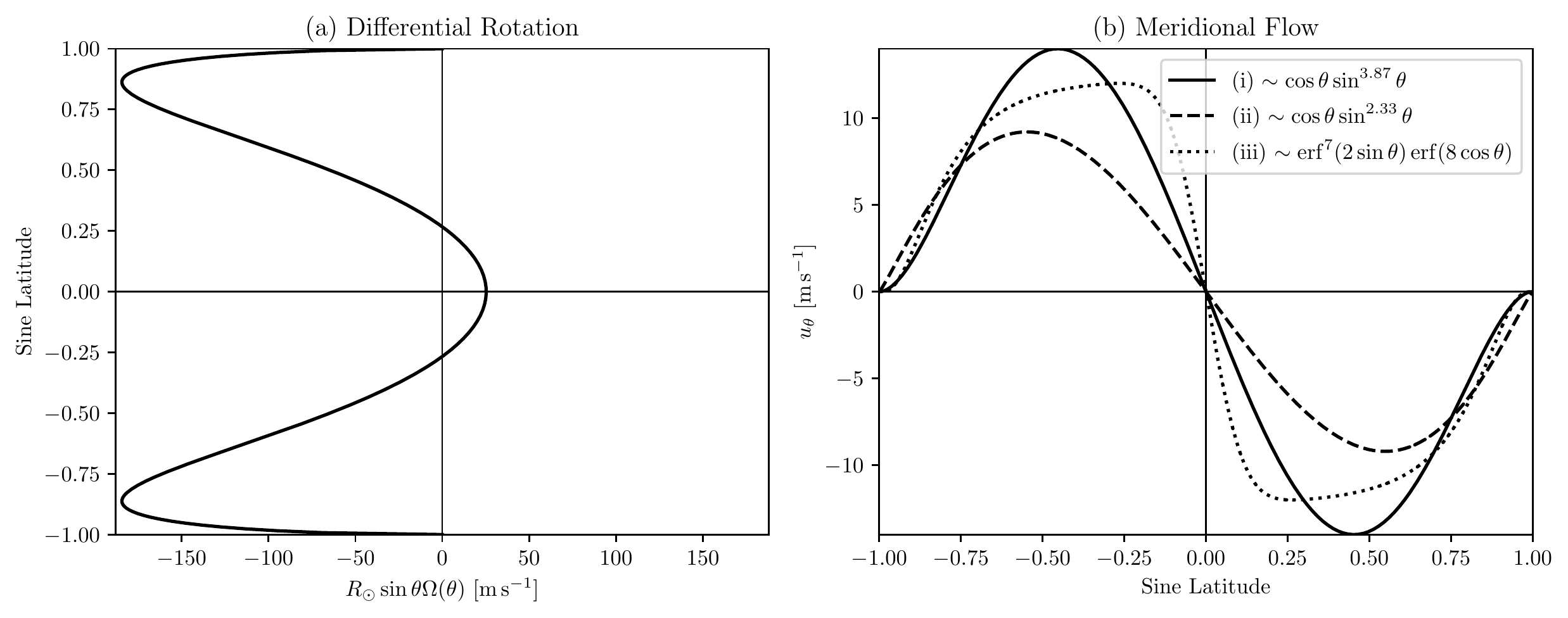}
    \caption{Velocity profile of differential rotation in the Carrington frame (a), and some example optimized profiles of meridional flow velocity (b), including (i) the simulation shown in Figure \ref{fig:paper_opt_bfly}; (ii) the Cycle 21 simulation of \cite{Whitbread2017}; and (iii) the Cycle 21 simulation of \cite{Lemerle2015}.  The corresponding values of $\Delta_u$ are (i) $0.7\times 10^{-7}\,\mathrm{s}^{-1}$, (ii) $0.4\times 10^{-7}\,\mathrm{s}^{-1}$, and (iii) $1.6\times 10^{-7}\,\mathrm{s}^{-1}$.}
    \label{fig:flow_profiles}
\end{figure}

The solar surface differential rotation is well constrained observationally \edit{\citep[see, e.g.][]{Beck2000}} and usually treated as a fixed constraint. Typically, SFT models use a steady axisymmetric angular velocity profile such as 
\begin{equation}
    \Omega(\theta) = 0.18 - 2.396\cos^2\theta - 1.787\cos^4\theta \quad \left[^\circ\,\mathrm{day}^{-1}\right]
    \label{eqn:omega}
\end{equation}
as determined by \cite{Snodgrass1990}. The constant term here is written in the Carrington frame that is usually adopted for SFT simulations. The resulting velocity profile is shown in Figure \ref{fig:flow_profiles}(a). As mentioned above, the differential rotation affects only the non-axisymmetric component of $B_r$, not the axisymmetric component $\langle B_r\rangle$, and will not be discussed further.

\subsection{Meridional Flow} \label{sec:merid}

\begin{figure}
    \centering
    \includegraphics[width=\textwidth]{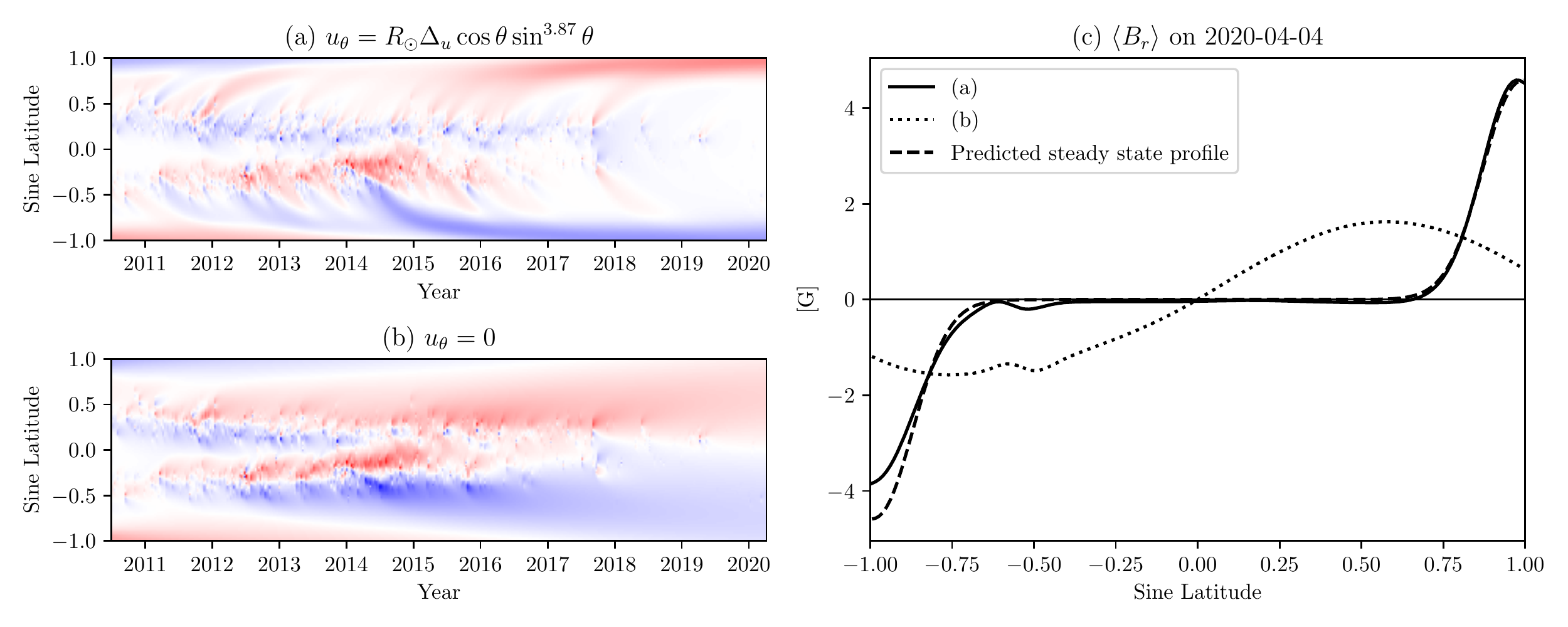}
    \caption{Effect of meridional flow in the simulation from Figure \ref{fig:paper_opt_bfly}, showing latitude-time plots of $\langle B_r\rangle$ when the flow is included (a) or omitted (b). Panel (c) shows the latitudinal profiles of $\langle B_r\rangle$ at the end of the simulation. The dashed curve shows the (near) steady-state profile \eqref{eqn:steady} for the case with flow. \citep[After Figure 3 of  ][]{Sheeley2005}.}
    \label{fig:paper_noflow}
\end{figure}

Although the only large-scale flow included by \cite{Leighton1964} was the differential rotation, it became clear from subsequent investigation of the SFT model that adding a meridional flow gives more realistic magnetic flux distributions \citep{DeVore1984}. In particular, a poleward flow is needed in order to concentrate the magnetic field into polar caps at the end of the solar cycle -- compare Figures \ref{fig:paper_noflow}(a) and (b). Otherwise, once $B_r$ has become approximately axisymmetric it will tend to the slowest decaying ($\ell=1$) eigenmode of the diffusion operator, which is the dipole $B_r\sim\cos\theta$. (A pure dipole is not seen in Figure \ref{fig:paper_noflow}c because it requires a few more years: the decay time for the next higher mode, $l=2$, is $R_\odot^2/[\eta l(l+1)] \approx 6\,\mathrm{yr}$.)

Observational evidence now clearly supports the existence of a surface meridional flow \citep{Hanasoge2022} although it is much slower than the differential rotation and potentially more variable. As such, different modellers have used different flow profiles. \edit{Typical examples have a single peak in each hemisphere, but vary in their latitudinal profiles. Figure \ref{fig:flow_profiles}(b) illustrates three profiles: (i) and (ii) come from the simple two-parameter family
\begin{equation}
    u_\theta(\theta) = \edit{-}R_\odot\Delta_u\cos\theta\sin^p\theta,
    \label{eqn:uthex}
\end{equation}
where $\Delta_u$ is the flow divergence at the equator, and larger values of $p$ lead to flows more concentrated near the equator (the speed peaks at $\cos\theta = \pm (1+p)^{-1/2}$). Profile (iii) in Figure \ref{fig:flow_profiles}(b) has the more complex form
\begin{equation}
    u_\theta(\theta) = \edit{-\frac{\sqrt{\pi}R_\odot\Delta_u}{2w\,\mathrm{erf}^q(\nu)}}\,\mathrm{erf}^q(\nu\sin\theta)\,\edit{\mathrm{erf}}(w\cos\theta),
    \label{eqn:uthlem}
\end{equation}
which allows the gradient to be concentrated nearer to the equator \citep[see][]{Wang2017}.
}

It is non-trivial to determine the precise eigenmodes of equation \eqref{eqn:sft1d} when meridional flow is included \citep{DeVore1987}, even with a simple flow profile such as \eqref{eqn:uthex}. However, one can determine a useful approximation by seeking a perfectly axisymmetric steady state $B_r(\theta)$ that balances the poleward advection with diffusion. For example, for the flow profile \eqref{eqn:uthex}, equation \eqref{eqn:sft1d} can be solved in an individual hemisphere to give the steady state solution
\begin{equation}
    B_r(\theta) = B_r(0)\exp\left[-\frac{\mathrm{Rm}_0\sin^{1+p}\theta}{(1+p)}\right].
    \label{eqn:steady}
\end{equation}
Here $\mathrm{Rm}_0 = R_\odot^2\Delta_u/\eta$, which is the magnetic Reynolds number $\mathrm{Rm}$ from \eqref{eqn:Lambda} with the specific choice $U_0 = R_\odot\Delta_u$, highlighting explicitly the dependence of the solution on the magnetic Reynolds number. The amplitude $B_r(0)$ will depend on the initial condition and source term $S$ and cannot be determined directly. The solution \eqref{eqn:steady} can only be an approximation to the slowest-decaying eigenfunction because it is necessarily non-zero at the equator, and will therefore generate a discontinuity at the equator when applied in both hemispheres with opposite sign. However, this discontinuity is small for typical values of $\mathrm{Rm}_0$ and will lead to diffusive cancellation only on a timescale much longer than the solar cycle \citep[cf.][]{Cameron2010}. Indeed, Figure \ref{fig:paper_noflow}(c) shows that \eqref{eqn:steady} gives an excellent approximation to the latitudinal $B_r$ profile at the end of the example simulation in Figure \ref{fig:paper_noflow}(a), particularly in the Northern hemisphere. (In the Southern hemisphere there is a remnant active region at low latitude that modifies the profile.) This simulation used $\eta=425\,\mathrm{km}^2\mathrm{s}^{-1}$, $p=3.87$, $\Delta_u=6.9\times 10^{-8}\,\mathrm{s}^{-1}$, and consequently $\mathrm{Rm}_0\approx 79$.

\subsection{Parameter Optimization}

The \edit{primary flow parameters to choose are the meridional flow profile $u_\theta(\theta)$ and} the diffusivity coefficient $\eta$. The basic effects of varying these parameters were investigated in the 1980s \citep{DeVore1984,Wang1989}. A more systematic parameter study was published by \cite{Baumann2004}, who explored the results of varying both $\eta$ and the meridional flow amplitude (in addition to properties of the source term), albeit varying only one parameter at a time and not the shape of the meridional flow profile.

More recent studies have explored the parameter space more widely, and have also attempted to optimize the parameters directly against synoptic magnetogram observations. The two most general studies are \cite{Lemerle2015} and \cite{Whitbread2017}, who both allow the strength and shape of $u_\theta(\theta)$ to vary, in addition to $\eta$. 
For $u_\theta$, \cite{Whitbread2017} allowed for profiles of the form \eqref{eqn:uthex}, whereas \cite{Lemerle2015} allow for the more general (but still single-peaked) \edit{form \eqref{eqn:uthlem}.} The optimal profiles from both studies for data from Cycle 21 are shown in Figure \ref{fig:flow_profiles}.
At present, it is not possible to select confidently between these solutions using observations, though helioseismic measurements of the plasma flow suggest equatorial slopes $\Delta_u$ in the range $[0.6-1.2]\times 10^{-7}\,\mathrm{s}^{-1}$ -- somewhere between profiles (i) and (iii) in Figure \ref{fig:flow_profiles}. Measurements based on magnetic feature tracking give lower equatorial slopes more like that of profile (ii), but it has been suggested that these are contaminated by supergranular diffusion \citep{Dikpati2010,Wang2017}. A recent list of observations is given in \citet{Jiang2022}.


Both \edit{\cite{Lemerle2015} and \cite{Whitbread2017}} used the same genetic optimization algorithm, PIKAIA \citep{Charbonneau1995}. \edit{These two studies} differed in their chosen goodness-of-fit functions, although both were ultimately derived from comparing to observed $B_r(\theta,\phi)$ maps. \cite{Whitbread2017} gave more weight to lower latitudes (where magnetogram observations are more reliable), whereas \cite{Lemerle2015} gave additional weight to the mid-latitude ``transport regions'' (because they represent the result of the model evolution rather than only the active region emergence) and to the axial dipole strength. At the other extreme, \edit{a further parameter study by \cite{Petrovay2019} focused only on optimizing the high latitude (polar) field, albeit in the 1D model}. This study used a synthetic (averaged) source term and fitted to average cycle properties from Wilcox Observatory polar field measurements, such as reversal time or width of the polar cap.

\begin{figure}
    \centering
    \includegraphics[width=0.8\textwidth]{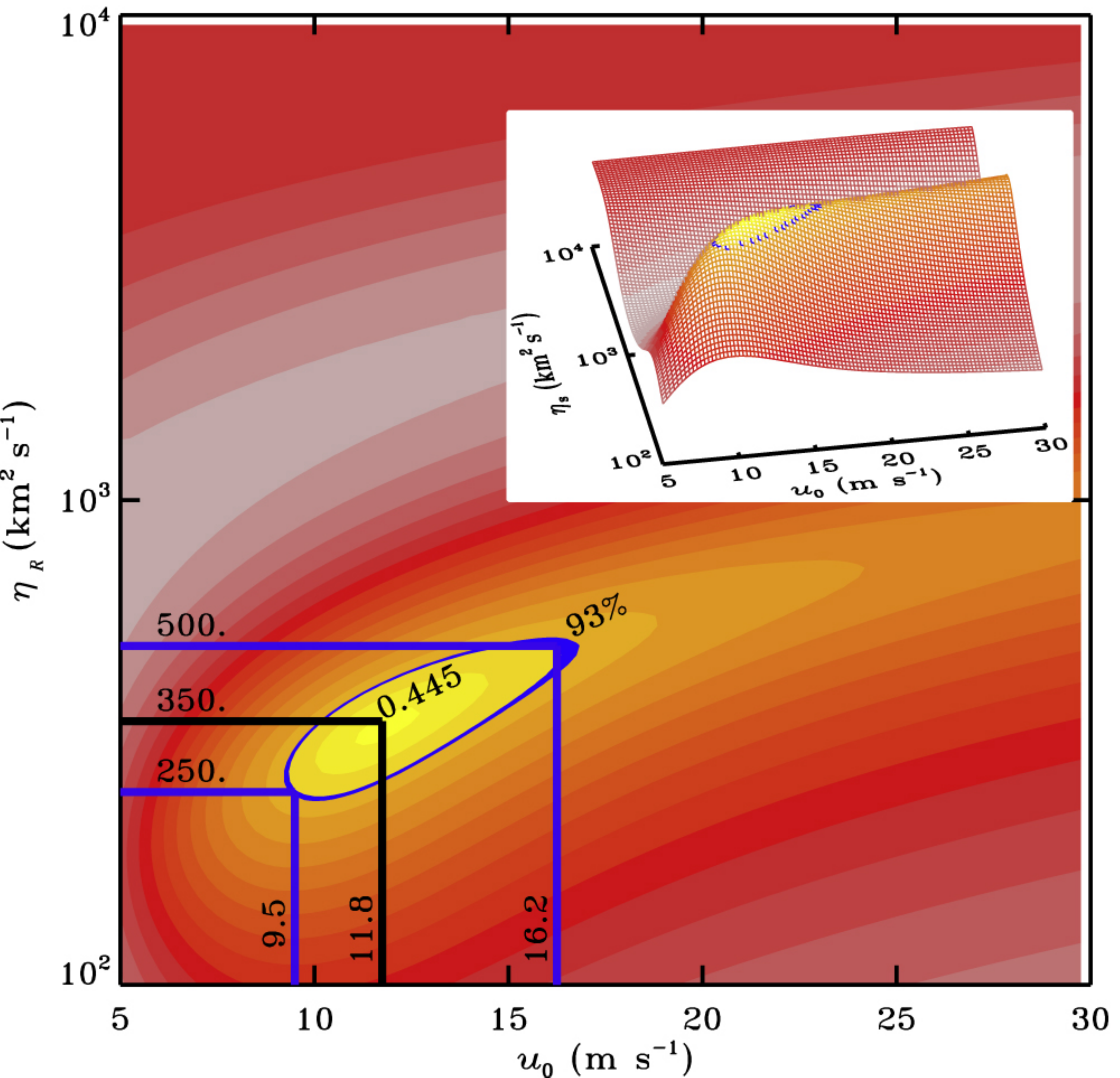}
    \caption{Fitness function $\chi^{-2}$ as a function of meridional flow amplitude $u_0=\max_\theta\lvert u_\theta\rvert$ (horizontal axis) and diffusivity $\eta_{\rm R}\equiv \eta$ (vertical axis), from the optimization study of \cite{Lemerle2015}. Black lines show the optimum value and blue lines the limit of the acceptable region ($\chi^{-2}\geq 93\%\chi^{-2}_{\rm max}$). (\copyright\,AAS. Reproduced with permission. Original article: \url{http://dx.doi.org/10.1088/0004-637X/810/1/78})}
    \label{fig:lemerle2015}
\end{figure}

A robust finding in these optimization studies is a degeneracy between $\eta$ and the amplitude of $u_\theta$. This is illustrated by Figure \ref{fig:lemerle2015}, which shows that there is a long ridge of near-optimal solutions in parameter space. Increasing both parameters together tends to lead to a equally (or nearly equally) well-matched solution, perhaps explaining why different groups have been able to use quite different values of $\eta$ -- for example, \cite{Cameron2010} use $\eta=250\,\mathrm{km}^2\mathrm{s}^{-1}$ as their standard value whereas the simulation in Figure \ref{fig:paper_opt_bfly} used $\eta=425\,\mathrm{km}^2\,\mathrm{s}^{-1}$.  This degeneracy makes sense given the appearance of the magnetic Reynolds number $\mathrm{Rm}$ in equation \eqref{eqn:sftnd}, which is essentially the ratio of $\eta$ to $\lvert u_\theta\rvert$. It means that SFT simulations can not be used to constrain both the meridional flow and diffusion from magnetogram observations alone.

When optimizing the model individually for different solar cycles, \cite{Whitbread2017} found some cycle-to-cycle variation in the optimal speeds and diffusivities. This is understandable given the phenomenological nature of the model (to be discussed further in Section \ref{sec:justification}).
Indeed, when simulating multiple cycles, \cite{Wang2002} had previously varied the meridional flow speed from cycle to cycle so as to avoid unrealistic drift of the polar field over time. On the other hand, other authors have avoided this problem by varying instead the tilts of emerging active regions \citep[][see also Section \ref{sec:modsource}]{Cameron2010}, or adding an addition decay term (to be discussed in Section \ref{sec:justification}).
In reality it is likely that the effective mean-field meridional flow varies even over the course of a single Solar Cycle (see Section \ref{sec:beyond}). Interestingly, \citet{hung2017} have shown -- in the context of a flux-transport (interior) dynamo model -- that a time-dependent meridional flow may be recovered from surface magnetic data through variational data-assimilation, and in future this approach could also be \edit{applied to} SFT.

\section{The Source Term}\label{sec:source}

The magnetic flux patterns in the SFT model are determined in large part by the source term $S(\theta,\phi,t)$, which -- in the classical mean-field model -- represents the emergence of new macroscopic active regions on the solar surface.
Since the classical SFT equation \eqref{eqn:sft} is linear in $B_r$, the solution is a superposition of solutions for each individual active region, so it is insightful to consider the evolution of one of these regions in isolation. Since most SFT simulations follow the evolution for periods of years, it is usual to emerge each active region instantaneously in time, so that
\begin{equation}
    S(\theta,\phi, t) = \sum_{i} B_r^{(i)}(\theta,\phi)\delta(t-t^{(i)}),
\end{equation}
where $B_r^{(i)}(\theta,\phi)$ is the magnetic field of an individual active region emerging at $t=t^{(i)}$.

\begin{figure}
    \centering
    \includegraphics[width=0.5\textwidth]{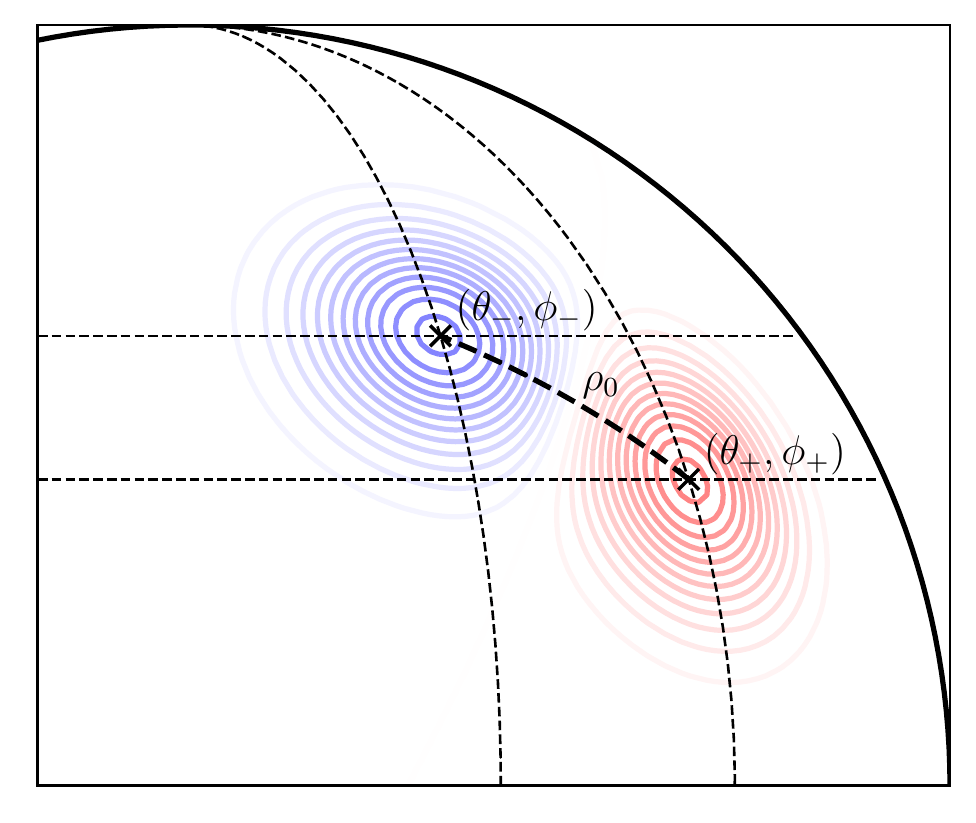}
    \caption{Positive and negative contours of $B_r$ for a BMR of the \cite{vanBallegooijen1998} form \eqref{eqn:bmrvb}. The size is exaggerated ($\rho_0=25^\circ$) compared to a real active region. This example follows Joy's Law in that the leading (rightmost) polarity is closest to the equator.}
    \label{fig:vb-bmr}
\end{figure}

Traditionally, SFT models treat each active region as a bipolar magnetic region (BMR). Figure \ref{fig:vb-bmr} shows the shape used by \cite{vanBallegooijen1998}, with circular flux patches centred on the poles $(\theta_-,\phi_-)$ and $(\theta_+,\phi_+)$ and having the form
\begin{equation}
    B_r(\theta,\phi) = B_0\left\{\exp\left[-\frac{2(1 - \cos\beta_+)}{(b\rho_0)^2}\right] - \exp\left[-\frac{2(1 - \cos\beta_-)}{(b\rho_0)^2}\right]\right\},
    \label{eqn:bmrvb}
\end{equation}
where 
\begin{align}
    \cos\beta_\pm &= \cos\theta_\pm\cos\theta + \sin\theta_\pm\sin\theta\cos(\phi-\phi_\pm),\\
    \cos\rho_0 &= \cos\theta_+\cos\theta_- + \sin\theta_+\sin\theta_-\cos(\phi_+-\phi_-). \label{eqn:rho0}
\end{align}
Thus $\beta_\pm(\theta,\phi)$ denote the heliocentric angles from each pole, and $\rho_0$ the heliocentric angle between them. \cite{vanBallegooijen1998} \edit{took $b=0.4$.} For some purposes, one can approximate \eqref{eqn:bmrvb} with a pair of Dirac-delta sources,
\begin{equation}
    B_r(\theta,\phi) = \frac{\Phi_0}{ R_\odot^2\sin\theta}\Big[\delta(\theta-\theta_+)\delta(\phi-\phi_+) - \delta(\theta-\theta_-)\delta(\phi-\phi_-)\Big],
    \label{eqn:bmrdelta}
\end{equation}
\edit{where $\Phi_0$ gives the flux of each polarity, defined assuming flux balance as
\begin{equation}
\Phi_0 = \frac{R_\odot^2}{2}\int_S\lvert B_r\rvert\sin\theta\,\mathrm{d}\theta\,\mathrm{d}\phi.
\end{equation}
}

Although the precise chosen shape for BMRs varies between implementations \cite[see][for another variation]{Yeates2020}, the key properties are the magnetic flux, $\Phi_0$, and pole locations, $(\theta_-,\phi_-)$ and $(\theta_+,\phi_+)$. The latter may equivalently be specified by giving the coordinates of the BMR centre $(\theta_0,\phi_0)$ along with the separation $\rho_0$ as in \eqref{eqn:rho0} and tilt angle $\gamma_0$, typically defined by
\begin{equation}
    \tan\gamma_0 = \frac{\theta_+ - \theta_-}{\sin\theta_0(\phi_+ - \phi_-)}.
    \label{eqn:tantilt}
\end{equation}
Together these BMR properties determine both the short-term and long-term evolution of the region.

\begin{figure}
    \centering
    \includegraphics[width=\textwidth]{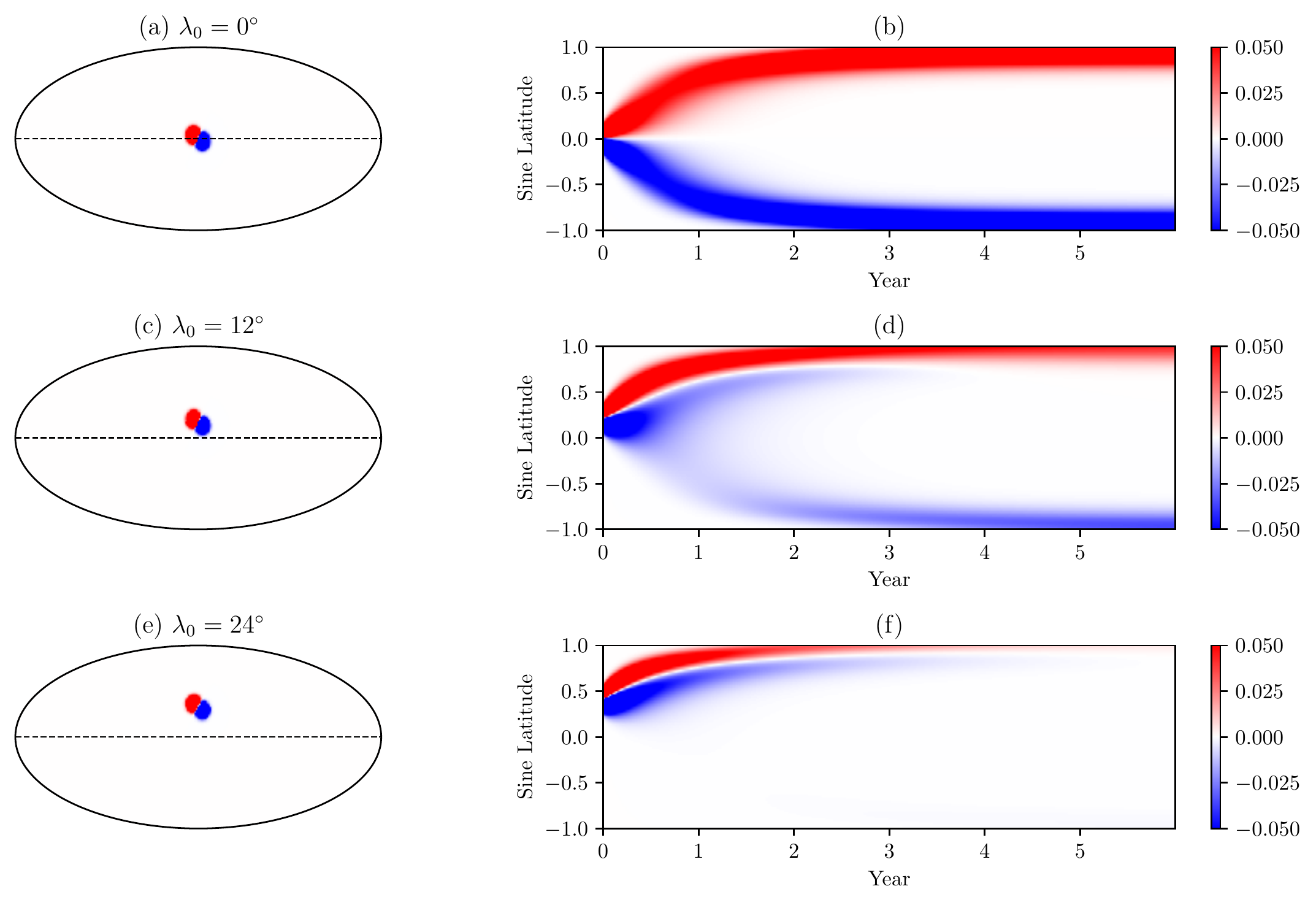}
    \caption{Long term evolution of three identical BMRs emerged at different latitudes ($\lambda_0 = \pi/2 - \theta_0$) in the SFT model. Left column shows the initial BMRs and right column the time-latitude plot of $\langle B_r\rangle$ in each case.}
    \label{fig:bmr-evo}
\end{figure}

After a new region emerges \edit{in the model}, much of its magnetic flux cancels by supergranular diffusion. \edit{This} models the observed process of flux cancellation at the polarity inversion line (PIL) between the positive and negative polarities. This cancellation rate is enhanced as the region is sheared by differential rotation and the PIL lengthened. 
On short-timescales (days) it is possible to approximate the solar surface as a Cartesian plane. Assuming a linear shear flow profile for the differential rotation, Lagrangian variables can be used to solve the Cartesian form of \eqref{eqn:sft} for the exact evolution $B_r(\theta,\phi,t)$ of a tilted BMR (we will see an example in Section \ref{sec:amp}). On longer timescales, it is necessary to follow the evolution numerically.

It takes approximately 2 years for the non-axisymmetric component of $B_r$ to cancel completely \citep{Wang1991}. Whether or not any axisymmetric $B_r$ remains on a longer timescale depends on how much flux of one polarity escapes across the equator so that the two polarities are pushed to opposite poles by the meridional flow. An untilted region will send both polarities equally to each pole and so leave no asymptotic contribution at the end of the solar cycle.
In a similar way, a (tilted) region that is nearer to the equator will produce a greater asymptotic contribution, because more flux escapes across the equator before being cancelled. This important effect is illustrated in Figure \ref{fig:bmr-evo}, where the same BMR is inserted at three different latitudes.

\subsection{Dipole Amplification Factor of a BMR} \label{sec:amp}

A common way to measure the end-of-cycle contribution of an individual BMR is through its axial dipole strength\edit{, which is the axisymmetric spherical harmonic coefficient of $B_r$ with lowest degree,}
\begin{equation}
    b_{1,0}(t) = \frac{3}{4\pi}\int_0^{2\pi}\int_0^\pi B_r\cos\theta\sin\theta\,\mathrm{d}\theta\,\mathrm{d}\phi = \frac{3}{2}\int_0^\pi\langle B_r\rangle\cos\theta\sin\theta\,\mathrm{d}\theta.
    \label{eqn:b10}
\end{equation}
By linearity of the classical SFT model, the total axial dipole strength will be the sum of the individual contributions from all of the active regions.

At the time of emergence, a BMR with the simple form \eqref{eqn:bmrdelta} has
\begin{align}
b_{1,0}(t_{\rm em}) &= \frac{3\Phi_0}{4\pi R_\odot^2}\int_0^\pi\Big[\delta(\theta-\theta_+) - \delta(\theta-\theta_-)\Big]\cos\theta\,\mathrm{d}\theta\\
    &=\frac{3\Phi_0}{4\pi R_\odot^2}\big(\cos\theta_+ - \cos\theta_-\big)\\
    &= \frac{3\Phi_0}{2\pi R_\odot^2}\sin\left(\frac{\theta_--\theta_+}{2}\right)\sin\left(\frac{\theta_++\theta_-}{2}\right)\\
    &\approx -\frac{3\Phi_0}{4\pi R_\odot^2}\rho_0\sin\gamma_0\sin\theta_0. \label{eqn:binit}
\end{align}
Here we have defined the central colatitude $\theta_0=(\theta_++\theta_-)/2$ and recognized that for tilt angle $\gamma_0$ and heliocentric angle $\rho_0$ between the poles, their latitudinal  separation is $(\theta_+-\theta_-)=\rho_0\sin\gamma_0$ (assuming $\theta_+>\theta_-$). Thus, as noted by \cite{Wang1991}, the axial dipole strength of a newly-emerged BMR depends on its flux, its latitudinal pole separation,  and the cosine of its emergence latitude.

\begin{figure}
    \centering
    \includegraphics[width=0.6\textwidth]{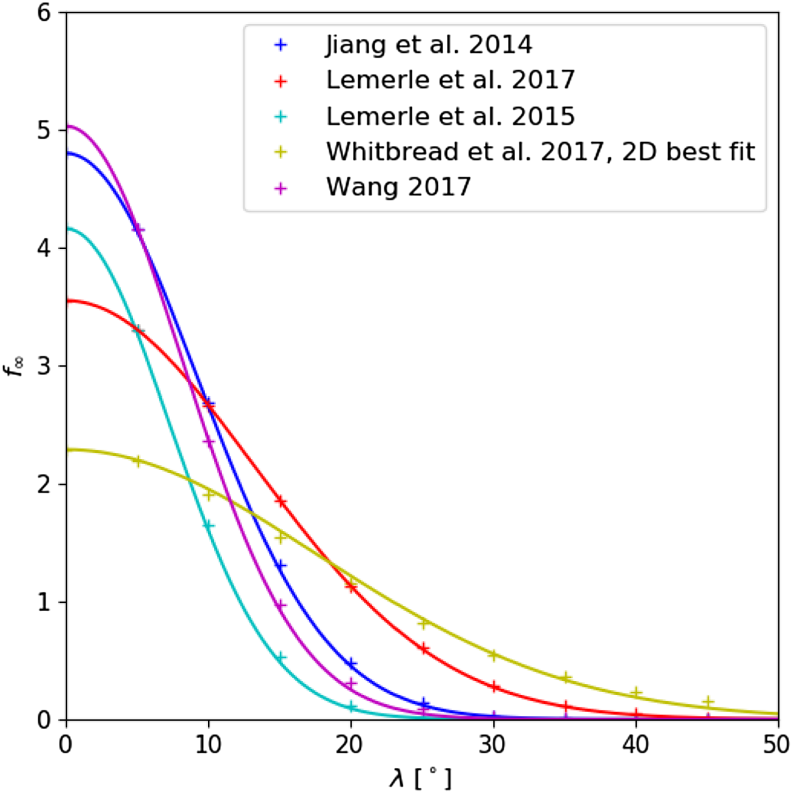}
    \caption{Latitude dependence of the dipole amplification factor for BMRs in different published SFT models. The solid lines show Gaussian fits. Reproduced from \cite{Petrovay2020}.}
    \label{fig:gaussian}
\end{figure}

Importantly, the axial dipole strength of a BMR can change under the ensuing SFT evolution: it will be amplified if the BMR emerged near the equator, or will decay if the BMR emerged far from the equator. It was first recognized by \cite{Jiang2014} that the ``dipole amplification factor''
\begin{equation}
        f_\infty = \lim_{t\to\infty}\frac{b_{1,0}(t)}{b_{1,0}(t_{\rm em})}
    \label{eqn:finf1}
\end{equation}
is well approximated by a Gaussian function of latitude, of the form
\begin{equation}
    f_\infty(\lambda_0) = A\,\exp\left(-\frac{\lambda_0^2}{2\lambda_R^2}\right),
    \label{eqn:finf}
\end{equation}
where \edit{$\lambda_0=\pi/2 - \theta_0$ is the central latitude of the BMR. (It is convenient to work in terms of latitude $\lambda=\pi/2 - \theta$ rather than colatitude $\theta$.)} Figure \ref{fig:gaussian} shows the functional form measured in several different numerical SFT models, where we note that both the amplitude $A$ and width $\lambda_R$ depend on the model. Once these parameters are known, equation \eqref{eqn:finf} -- coupled with the linearity of the SFT evolution equation \eqref{eqn:sft2d} or \eqref{eqn:sft1d} -- allows the net axial dipole strength at the end of a solar cycle to be determined algebraically just by adding up the contributions of the individual BMRs, without the need to solve the \edit{evolution equation.}

The interpretation of Figure \ref{fig:gaussian} is that only BMRs that emerge with latitude $\lvert\edit{\lambda_0}\rvert<\lambda_R$ will contribute to the global dipole moment at the end of the solar cycle. \cite{Petrovay2020} call $\lambda_R$ the ``dynamo effectivity range'', and give the following simple physical derivation. To give a lasting contribution, a BMR must be close enough to the equator that some of its leading-polarity flux is able to cross the equator by diffusion, in opposition to the meridional flow. The timescale for advective separation at the equator is $\Delta_u^{-1}$, where $\Delta_u=R_\odot^{-1}u_\theta'(\pi/2)$ is the equatorial divergence of $u_\theta(\theta)$. Equating this to the diffusion timescale  $(\lambda R_\odot)^2/\eta$ from latitude $\lambda$ to the equator suggests that
\begin{equation}
    \lambda_R \approx \sqrt{\frac{\eta}{R_\odot^2\Delta_u}} = \mathrm{Rm}_0^{-1/2}.
    \label{eqn:lambdaR}
\end{equation}
Note the reappearance of the magnetic Reynolds number from Section \ref{sec:merid}. \cite{Petrovay2020} computed $f_\infty(\lambda_0)$ for numerical solutions with several different $u_\theta$ profiles, and in most cases found that the Gaussian width $\lambda_R$ was indeed well approximated by $\mathrm{Rm}_0^{-1/2}$, the exception being a flow where $u_\theta$ peaks at a very low latitude compared to observations.

\cite{Petrovay2020} went further and derived \eqref{eqn:finf} analytically. \edit{(Readers not interested in the details may skip to Section \ref{sec:nbps}.)} The trick is to recognize that the final dipole moment -- once the $B_r$ distribution has become (near) axisymmetric -- will be proportional to the remaining net magnetic flux in each hemisphere. (There will also be a coefficient depending on the latitudinal profile of the near-steady state as in \eqref{eqn:steady}.) Because it is determined purely by flux crossing the equator, the evolution of the net hemispheric flux can be quite well approximated by a Cartesian SFT model near the equator, which has the advantage of being analytically tractable. Thus \cite{Petrovay2020} consider the ``low-latitude limit'' of \eqref{eqn:sft1d},
\begin{equation}
        \frac{\partial\langle B_r\rangle}{\partial t} + \frac{1}{R_\odot}\frac{\partial}{\partial\lambda}\big(u_\lambda\langle B_r\rangle\big) = \frac{\eta}{R_\odot^2}\frac{\partial^2\langle B_r\rangle}{\partial\lambda^2}.
    \label{eqn:sftcart}
\end{equation}
By choosing the linearised meridional flow $u_\lambda = R_\odot\Delta_u\lambda$, we can define the Lagrangian coordinate $\ell = \mathrm{e}^{-\Delta_u t}\lambda$ and new time variable $\tau = \left(1 - \mathrm{e}^{-2\Delta_u t}\right)/(2\mathrm{Rm}_0)$ to reduce Equation \eqref{eqn:sftcart} to a standard diffusion equation
\begin{equation}
        \frac{\mathrm{D}}{\mathrm{D}\tau}\left(\mathrm{e}^{\Delta_u t}\langle B_r\rangle\right) = \frac{\partial^2}{\partial\ell^2}\left(\mathrm{e}^{\Delta_u t}\langle B_r\rangle\right),
        \label{eqn:diff}
\end{equation}
where \edit{$\mathrm{D}/\mathrm{D}\tau$} denotes the partial derivative with $\ell$ kept constant rather than $\lambda$. Equation \eqref{eqn:diff} may be solved for a variety of initial conditions using standard techniques.

If the initial condition consists of a single (monopole) point source,
\begin{equation}
    \langle B_r\rangle(\lambda,0) = \frac{\Phi_0}{2\pi R_\odot^2}\delta(\lambda-\lambda_0),
\end{equation}
then solving \eqref{eqn:diff} gives
\begin{equation}
    \mathrm{e}^{\Delta_u t}\langle B_r\rangle = \frac{\Phi_0}{2\pi R_\odot^2\sqrt{4\pi\tau}}\exp\left(-\frac{(\ell - \lambda_0)^2}{4\tau}\right),
\end{equation}
which for large $t$ is approximately
\begin{equation}
    \langle B_r\rangle(\lambda, t) \sim \frac{\Phi_0\sqrt{\mathrm{Rm}_0}\mathrm{e}^{-\Delta_u t}}{2\pi R_\odot^2\sqrt{2\pi}}\exp\left(\frac{-\mathrm{Rm}_0\left(\mathrm{e}^{-\Delta_u t}\lambda - \lambda_0\right)^2}{2}\right).
    \label{eqn:larget}
\end{equation}
In the approximation \eqref{eqn:larget}, the flux difference between the hemispheres is
\begin{align}
    \Phi_{\rm N} - \Phi_{\rm S} = 2\pi R_\odot^2\left(\int_0^\infty\langle B_r\rangle\,\mathrm{d}\lambda - \int_{-\infty}^0\langle B_r\rangle\,\mathrm{d}\lambda\right) =  \Phi_0\,\mathrm{erf}\left(\sqrt{\frac{\mathrm{Rm}_0}{2}}\lambda_0\right),
    \label{eqn:monofluxdiff}
\end{align}
valid for either sign of $\lambda_0$.

For a BMR we must combine two point sources as in \eqref{eqn:bmrdelta}, each contributing half of the flux $\Phi_0$, so 
\begin{align}
    \Phi_{\rm N} - \Phi_{\rm S} &= \frac{\Phi_0}{2}\left[\mathrm{erf}\left(\sqrt{\frac{\mathrm{Rm}_0}{2}}\lambda_+\right) - \mathrm{erf}\left(\sqrt{\frac{\mathrm{Rm}_0}{2}}\lambda_-\right)\right]\\
    &\approx \frac{\Phi_0\sqrt{\mathrm{Rm}_0}(\lambda_+ - \lambda_-)}{\sqrt{2\pi}}\exp\left(-\frac{\mathrm{Rm}_0\lambda_0^2}{2}\right),
\end{align}
where we recognize the finite difference as an approximation of the derivative at $\lambda_0 = (\lambda_-+\lambda_+)/2$. We therefore expect that, to a good approximation, $b_{1,0}(t) \to a(\Phi_{\rm N} - \Phi_{\rm S})/R_\odot^2$ as $t\to\infty$, for some constant $a$ that depends on the (normalized) shape of the steady $B_r$ profile (thus only on $u_\theta$ and $D$). At the initial time, Equation \eqref{eqn:binit} gives $b_{1,0}(0) = 3\Phi_0(\lambda_+ - \lambda_-)\cos\lambda_0/(4\pi R_\odot^2)$. Approximating $\cos\lambda_0\approx 1$, the ratio is therefore
\begin{equation}
    f_\infty \approx \frac{a\sqrt{8\pi\mathrm{Rm}_0}}{3}\exp\left(-\frac{\mathrm{Rm}_0\lambda_0^2}{2}\right).
    \label{eqn:finfanal}
\end{equation}
Thus we recover \eqref{eqn:finf} with $\lambda_R = \mathrm{Rm}_0^{-1/2}$ as claimed. Moreover, for a known asymptotic profile of $B_r(\theta)$, we can determine $a$ and hence also predict the amplitude $A$. 

\subsection{Non-Bipolar Source Regions} \label{sec:nbps}

\begin{figure}
    \centering
    \includegraphics[width=\textwidth]{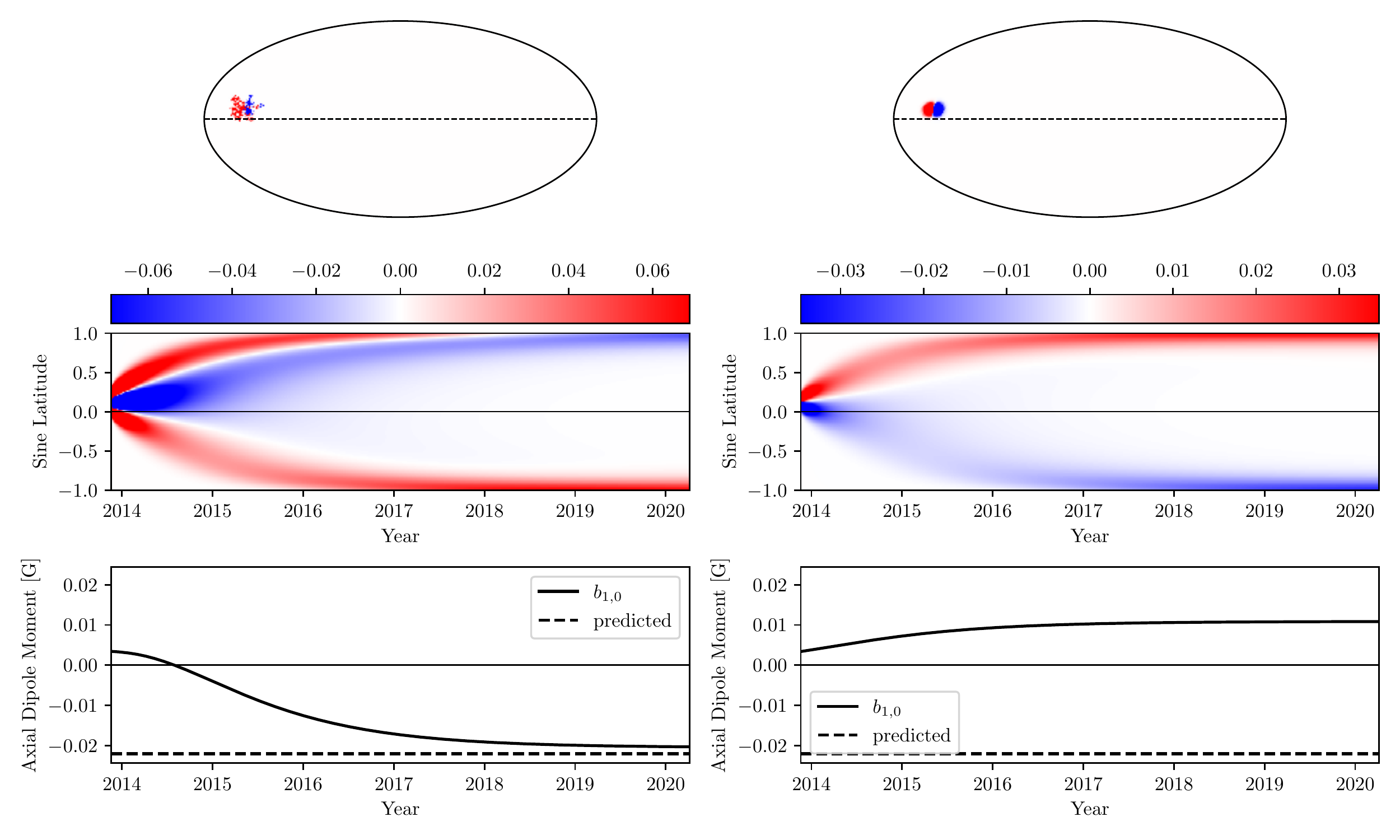}
    \caption{Evolution of an active region with asymmetric bipolar shape, taken from the simulation in Figure \ref{fig:paper_opt_bfly}. The left column shows the region with its \edit{observed} shape, with second row showing $\langle B_r\rangle$ and third row the axial dipole strength $b_{1,0}$. The right column shows the evolution of an ``equivalent'' symmetric BMR having the same initial flux and $b_{1,0}$. The dashed line shows the final $b_{1,0}$ predicted by equation \eqref{eqn:greens} using the observed magnetogram.}
    \label{fig:non-bmr}
\end{figure}

Real solar active regions cannot always be represented as simple, symmetric BMRs. Even a region with two polarities will be effectively ``multipolar'' if the polarities are asymmetric in shape, and this will modify the evolution of $b_{1,0}$ compared to a symmetric BMR. This was investigated by \cite{Iijima2019}, who ran SFT simulations with Gaussian BMRs of the form \eqref{eqn:bmrvb}, but where the leading polarity has a narrower width than the following polarity (controlled by the $b$ parameter in \eqref{eqn:bmrvb}). When calibrated to the observed level of sunspot area asymmetry, their SFT simulation gave a more realistic evolution of both $b_{1,0}$ and the magnetic butterfly diagram, as compared to a reference simulation with equally-sized polarities. In particular, they noted that a wider following polarity leads to more following polarity flux crossing the equator, cancelling some of the trans-equatorial leading polarity flux and weakening the asymptotic contribution of the region. Similarly, \cite{Wang2021} found for asymmetric BMRs with more diffuse following polarity, that $f_\infty$ is systematically reduced (see their Figure 4).
As an illustration, Figure \ref{fig:non-bmr} shows an example of the SFT evolution for an asymmetric region \edit{inserted directly with its observed shape}; in this case, the effect is sufficiently extreme to reverse the sign of $b_{1,0}$ altogether compared to a symmetric BMR. 

\cite{Jiang2019} considered the SFT evolution of a more complex ``$\delta$-type'' flux distribution. They showed that $b_{1,0}$   changed sign during the SFT evolution, ending up with a completely different end-of-cycle contribution than would be expected for a BMR emerging at the same latitude with the same flux and same initial $b_{1,0}$. \cite{Wang2021} showed further that the dipole amplification $f_\infty$ is no longer a simple function of emergence latitude for such complex regions. However, the net effect of all of the real complex and asymmetric regions seems to be a reduction in the net end-of-cycle dipole, at least for Cycle 24. Evidence for this comes from \cite{Yeates2020}, who compared an SFT simulation of that cycle where all active regions emerged with their observed flux distributions to a simulation where they were all approximated by symmetric BMRs with the same flux and initial $b_{1,0}$. The net $b_{1,0}$ at the end of the cycle was overestimated by 24\% when the regions were modelled with BMRs.

For predicting the dipole contributions of more complex regions, \cite{Wang2021} showed that \eqref{eqn:finfanal} can be generalized to regions with non-bipolar shapes, by treating them as a superposition of point sources. In particular, for an active region with initial flux distribution $B_r(\theta,\phi,0)$, combining the hemispheric flux differences \eqref{eqn:monofluxdiff} predicts that the axial dipole strength at the end of the cycle would be
\begin{equation}
    \lim_{t\to\infty}b_{1,0}(t) \approx \frac{a}{R_\odot^2}\int_SB_r(\theta,\phi,0)\,\mathrm{erf}\left[\sqrt{\frac{\mathrm{Rm}_0}{2}}\left(\frac{\pi}{2} - \theta\right)\right]\sin\theta\,\mathrm{d}\theta\mathrm{d}\phi,
    \label{eqn:greens}
\end{equation}
where $a$ is the coefficient in the relation $b_{1,0}\approx a(\Phi_{\rm N}-\Phi_{\rm S})/R_\odot^2$. \cite{Wang2021} verified this prediction against SFT simulations for 84 regions during Cycle 24. It gives an accurate prediction for the region in Figure \ref{fig:non-bmr}.

\subsection{Modelling the Source Term} \label{sec:modsource}

It is not always viable to use observations of real individual active regions to construct the source term. This situation arises when working with historical data, when running SFT models into the future for forecasting purposes, or just in conceptual simulations studying the underlying physics. In such cases the source term $S(\theta,\phi,t)$ needs to be modelled, either as a smooth function \citep[\textit{e.g.},][]{Cameron2007, Petrovay2019} or as random realizations of active regions drawn from a statistical distribution \citep[\textit{e.g.},][]{Schrijver2001,Mackay2002b,Baumann2004, Jiang2018, YMWang2021}.

\begin{figure}
    \centering
    \includegraphics[width=\textwidth]{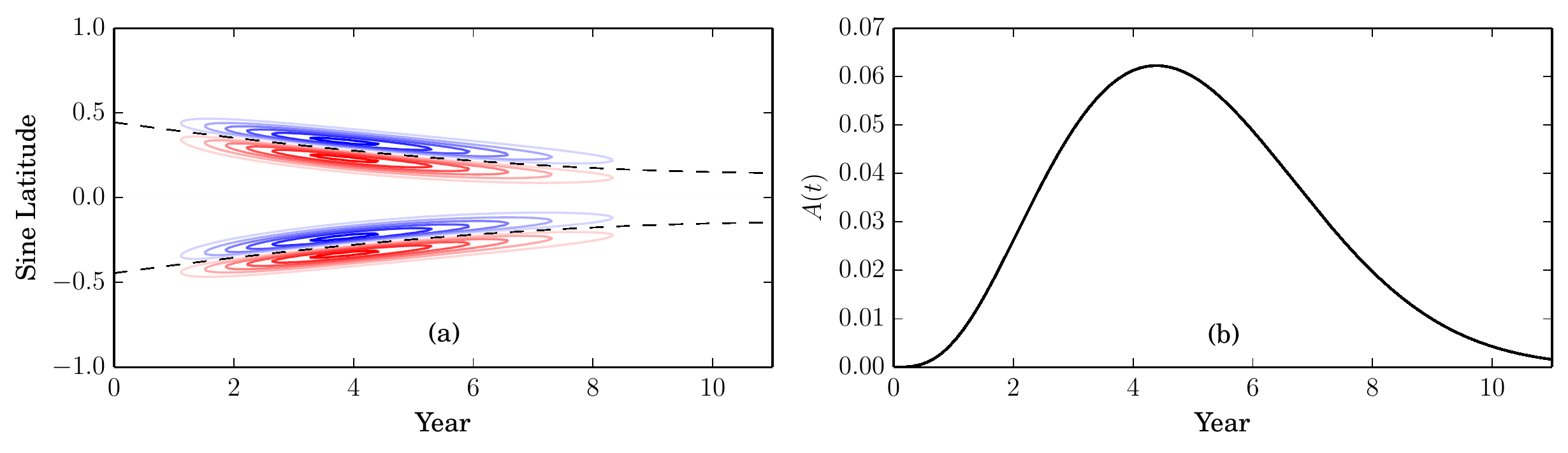}
    \caption{The smooth source term $\langle S\rangle(\lambda,t)$ used by \cite{Petrovay2019}. In (a), red/blue contours show $\langle S\rangle$, and
    dashed lines indicate $\pm \sin[\lambda_0(t)]$ from \eqref{eqn:cyccent}. Panel (b) shows the overall cycle shape $A(t)$ from \eqref{eqn:cycshape} with  $a=0.00185$, $b=4.058$, $c=0.71$.}
    \label{fig:smooth-source}
\end{figure}

The smooth function approach has primarily been used in the 1D SFT model, \eqref{eqn:sft1d}, for example by \citet{Petrovay2019}, who use a pair of flux rings in each hemisphere, shown in Figure \ref{fig:smooth-source}(a) and given by
\begin{align}
    \langle S\rangle(\lambda, t) &= (-1)^n A(t)\left\{\exp\left(-\frac{[\lambda-\lambda_+(t)]^2}{2\delta_\lambda^2}\right)
    - \exp\left(-\frac{[\lambda-\lambda_-(t)]^2}{2\delta_\lambda^2}\right)\right.\nonumber\\
    &\qquad+ \left.\exp\left(-\frac{[\lambda+\lambda_+(t)]^2}{2\delta_\lambda^2}\right)
    - \exp\left(-\frac{[\lambda+\lambda_-(t)]^2}{2\delta_\lambda^2}\right)\right\}.
    \label{eqn:petsource}
\end{align}
This model incorporates a number of observed solar cycle features:
\begin{enumerate}
    \item[(i)] All polarities alternate according to the solar cycle number, $n$.
    \item[(ii)] The cycle has an asymmetrical shape in time, shown in Figure \ref{fig:smooth-source}(b) and given by the \cite{Hathaway1994} observed fit 
        \begin{equation}
            A(t) = a(t-t_{\rm min})\left(\exp\left[\frac{(t-t_{\rm min})^2}{b^2}\right] - c\right)^{-1},
            \label{eqn:cycshape}
        \end{equation}
    where $t_{\rm min}$ is the start of the cycle.
    \item[(iii)] The centres $\pm\lambda_0$ of each pair of flux rings, \textit{i.e.} $\lambda_0 = (\lambda_+ + \lambda_-)/2$ shown by dashed lines in Figure \ref{fig:smooth-source}(a), migrate equatorward at the rate
        \begin{equation}
        \lambda_0(t) = 26.4 -  34.2\left(\frac{t}{T}\right) + 16.1  \left(\frac{t}{T}\right)^2\quad [^\circ]
        \label{eqn:cyccent}
        \end{equation}
    fitted empirically by \cite{Jiang2011}, where $T$ is the cycle length (11 years).
    \item[(iv)] The separation $\Delta_\lambda = \lambda_--\lambda_+$ decreases as $\lambda_0$ approaches the equator, according to
    \begin{equation}
        \Delta_\lambda(t) = 0.5\frac{\sin\lambda_0(t)}{\sin 20^\circ} \quad [^\circ].
        \label{eqn:joy}
    \end{equation}
    This models the longitude-averaged effect of the well-established Joy's Law \citep{vanDriel2015}, whereby BMRs emerging at lower latitude have (on average) smaller tilt angle $\lvert\gamma_0\rvert$, defined in \eqref{eqn:tantilt}.
\end{enumerate}

The statistical BMR approach is similar, except the functions above are treated as overall distributions from which discrete BMRs are chosen at random. For the longest historical simulations, which date back to 1700 \citep{Jiang2018, Wang2021}, the only observational input is the sunspot number time series -- equivalent to emergence rate, $A(t)$. For 20th Century simulations, data on the areas and locations of individual sunspot groups can be used \citep[\textit{e.g.},][]{Cameron2010}. However, even here the magnetic flux and tilt angle (equivalently axial dipole strength) must be chosen at random as they are not available observationally before the onset of routine magnetograms in the 1970s.

The tilt angle is problematic as Joy's Law, as modelled in \eqref{eqn:joy}, holds only for the mean, and there is known to be very significant scatter \citep[\textit{e.g.},][]{Wang1989b,Yeates2020}.
 Recent studies have shown that individual ``rogue'' active regions -- defined as those with dipole moments significantly different from Joy's Law expectation at their latitude -- can have a significant effect on the overall polar field at the end of the cycle \citep{Jiang2015, Nagy2017}. In light of \eqref{eqn:finfanal}, such rogue regions must typically emerge near to the equator, although their relative contribution depends on $\mathrm{Rm}_0$ and would be reduced if $\mathrm{Rm}_0$ were large. Nevertheless, simulations based on statistical source terms without individual dipole moment data should be treated with caution, particularly for prediction.
 
The widely-accepted $\alpha\Omega$ paradigm for the solar dynamo suggests a  ``self-consistent'' way to build a fully synthetic SFT model: set the amount flux emerging through the source term in cycle $n$ proportional to the axial dipole strength at the end of cycle $n-1$. \cite{Talafha2022} modified the one-dimensional model of \cite{Petrovay2019} to use such an approach. They used this model to  systematically study the impact of two possible nonlinearities in the source term: tilt quenching (where BMRs are less tilted in strong cycles) and latitude quenching (where BMRs emerge at higher latitudes in strong cycles). SFT simulations show that both effects act to reduce the axial dipole produced in strong cycles \citep{Cameron2010,Jiang2020}. They are both therefore possible saturation mechanisms to explain why the solar dynamo doesn't exhibit runaway exponential growth. \cite{Talafha2022} showed that the relative impact of tilt versus latitude quenching on the end-of-cycle axial dipole  depends primarily on the dynamo effectivity range $\lambda_R$ in equation \eqref{eqn:lambdaR}. In particular, for small $\lambda_R$, latitude quenching reduces the end-of-cycle dipole more than tilt quenching, and \textit{vice versa} for large $\lambda_R$. However, the amount of tilt and/or latitude quenching present on the real Sun remains under debate.

\section{Physical Justification}\label{sec:justification}

As introduced by \cite{Leighton1964}, the SFT model is purely phenomenological. But can equation \eqref{eqn:sft} be derived from known physical laws?  The relevant law governing the evolution of the large-scale magnetic field is the mean-field MHD (magnetohydrodynamic) induction equation,
\begin{equation}
    \frac{\partial B_r}{\partial t} = {\bf e}_r\cdot\nabla\times\big({\bf u}\times{\bf B} - \eta\nabla\times{\bf B}\big),
    \label{eqn:induc}
\end{equation}
where ${\bf u}$ is the plasma velocity and -- anticipating the form of \eqref{eqn:sft} -- we have made a simple approximation for the turbulent electromotive force of the form $-\eta\nabla\times{\bf B}$ \citep[cf.][]{McCloughan2002}. Thus $\eta$ represents turbulent diffusivity, not ohmic resistivity (which is negligible in the highly conducting photosphere). This assumption of a turbulent diffusivity is discussed further in Section \ref{sec:pumping} below.

Consider the first term in \eqref{eqn:induc}. Decomposing ${\bf u}={\bf u}_h + u_r{\bf e}_r$, where ${\bf e}_r\cdot{\bf u}_h=0$, and similarly ${\bf B}={\bf B}_h + B_r{\bf e}_r$, we can write
\begin{equation}
    {\bf e}_r\cdot\nabla\times\big({\bf u}\times{\bf B}\big) =  \nabla\cdot\big(u_r{\bf B}_h\big) - \nabla\cdot\big({\bf u}_hB_r\big).
\end{equation}
The last term is precisely the advection term in the SFT equation \eqref{eqn:sft}, while the term $\nabla\cdot\big(u_r{\bf B}_h\big)$ represents flux emergence, so corresponds to the source term $S$ in \eqref{eqn:sft}. Thus the SFT model is incorporating the correct advection terms.

Now consider the diffusion term in \eqref{eqn:induc}. For simplicity, we will assume that $\eta=\eta(r)$ only, in which case
\begin{align}
    -{\bf e}_r\cdot\nabla\times\big(\eta\nabla\times{\bf B}\big) &= \eta\nabla_h^2 B_r + R_\eta.
\end{align}
This has the diffusion term from \eqref{eqn:sft} plus an additional remainder term
\begin{align}
    R_\eta &= -\frac{\eta}{R_\odot}\nabla_h\cdot{\bf B} - \frac{\eta}{R_\odot\sin\theta}\frac{\partial}{\partial\theta}\left(\sin\theta\frac{\partial B_\theta}{\partial r}\right) - \frac{\eta}{R_\odot\sin\theta}\frac{\partial}{\partial\phi}\left(\frac{\partial B_\phi}{\partial r}\right).
    \label{eqn:extrar}
\end{align}
Using $\nabla\cdot{\bf B}=0$, this may be rewritten entirely in terms of $B_r$, simplifying to
\begin{align}
    R_\eta = \frac{\eta}{R_\odot^2}\frac{\partial^2}{\partial r^2}\big(r^2 B_r\big).
    \label{eqn:rdiff}
\end{align}
Thus in mean-field MHD there is an additional term representing the radial diffusion of magnetic flux that is missing from the original SFT equation \eqref{eqn:sft}. Physically, this incorporates the fact that the surface magnetic field is connected to the interior; for example, the decay of active regions can be slowed if they remain connected to deeper layers of the convection zone where the diffusivity is lower \citep{Wilson1990, Whitbread2019}.

\edit{One way to justify the classical SFT model is to assume that $B_\theta, B_\phi \approx 0$ in the near-surface region of the solar convection zone. It then follows from \eqref{eqn:extrar} that $R_\eta=0$. To some extent this is justified by vector magnetogram observations at the photosphere \citep[for a recent discussion, see][]{Virtanen2019}; for a theoretical argument, see \cite{vanBallegooijen2007}.
If this radial-field approximation is not made, then} self-consistent computation of the radial diffusion term $R_\eta$ would require simulation of the three-dimensional magnetic field in the solar convection zone. However, two approaches have been used to parametrize \eqref{eqn:rdiff} in SFT models without the need for three-dimensional simulations, and these will be considered next. 

\subsection{Exponential Decay Term} \label{sec:exp}

The most common parametrization for the radial diffusion term \eqref{eqn:rdiff} is to assume that $R_\eta\approx - B_r/\tau$, so that \eqref{eqn:sft} becomes
\begin{equation}
    \frac{\partial B_r}{\partial t} + \nabla_h\cdot\big({\bf u}_h B_r\big) = \eta\nabla_h^2 B_r - \frac{B_r}{\tau} + S.
\end{equation}
Multiplying by $\mathrm{e}^{t/\tau}$ shows that
\begin{equation}
    \frac{\partial}{\partial t}\big(\mathrm{e}^{t/\tau}B_r\big) + \nabla_h\cdot\big({\bf u}_h\mathrm{e}^{t/\tau}B_r\big) = \eta\nabla_h^2\big(\mathrm{e}^{t/\tau}B_r\big) + \mathrm{e}^{t/\tau}S.
\end{equation}
Thus if $B_r^\infty$ denotes the solution to the original equation \eqref{eqn:sft}, corresponding to $\tau\to\infty$, then the solution with finite $\tau$ but all other parameters the same is $B_r = \mathrm{e}^{-t/\tau}B_r^{\infty}$. In other words, the solution decays exponentially at uniform rate $\tau^{-1}$. For example, the dipole amplification factor \eqref{eqn:finfanal} for a BMR would become
\begin{equation}
    f_\infty \approx \frac{\sqrt{8\pi\mathrm{Rm}_0}}{3}\exp\left(-\frac{\mathrm{Rm}_0\lambda_0^2}{2}\right)\exp\left(-\frac{t}{\tau}\right),
\end{equation}
reflecting continuing decay of the magnetic field due to the new term.

\begin{figure}
    \centering
    \includegraphics[width=0.8\textwidth]{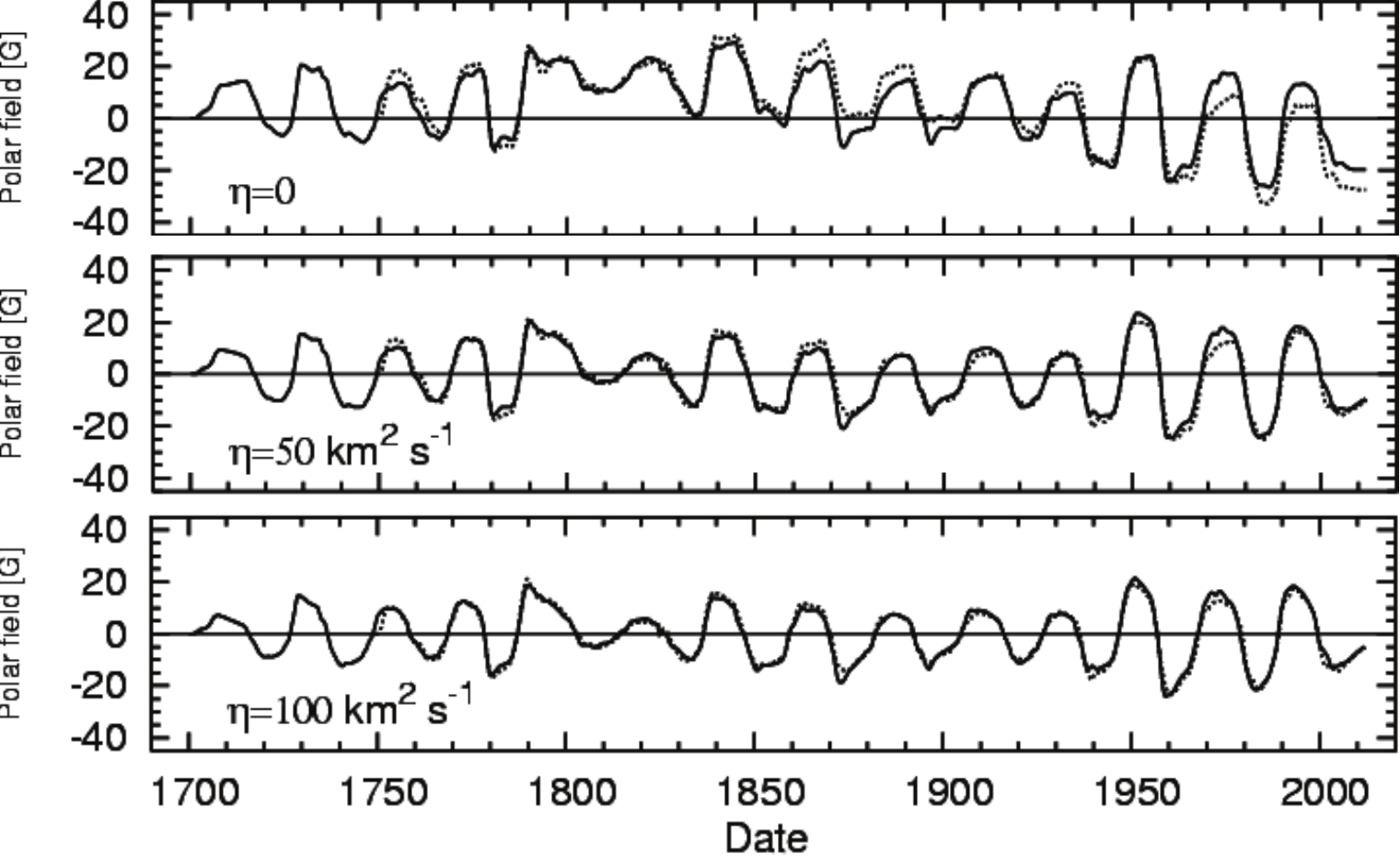}
    \caption{Application of the radial diffusion term to reduce spurious cycle-to-cycle memory in the SFT model, from \cite{Baumann2006}. The top row shows the north polar field (above $75^\circ$ latitude) in a simulation with no radial diffusion, while the middle and bottom rows show the same simulation with $\eta_0=50\,\mathrm{km}^2\mathrm{s}^{-1}$ and $100\,\mathrm{km}^2\mathrm{s}^{-1}$ according to the prescription in Section \ref{sec:baumann}. The simulation uses random emerging BMRs proportional to the observed sunspot numbers. The dashed line shows a simulation started in 1750, illustrating how the memory of the initial conditions persists. (Credit: Baumann, I., Schmitt, D. and Sch\"{u}ssler, M., A\&A, 446, 307-314, 2006, reproduced with permission \copyright\,ESO.)}
    \label{fig:decay}
\end{figure}

The first application of such a decay term was by \cite{Schrijver2002}, who motivated it not by consideration of radial diffusion but purely as a necessary addition to reduce the ``memory'' of the polar field (equivalently $b_{1,0}$) over multiple solar cycles. Without it, the varying amount of polar field production caused by the differing sunspot numbers in different cycles led to an unrealistic drift in the polar field over time, rather than the regular reversals that are observed. This drift is illustrated (for another SFT model) in the top panel of Figure \ref{fig:decay}.

The optimization studies discussed in Section \ref{sec:flows} have also looked for the optimum $\tau$ in shorter simulations where long-term memory is not an issue. With their simplified source term, \cite{Petrovay2019} found that a decay term (with $\tau$ in the range 5--10$\,\mathrm{yr}$) was essential, otherwise $b_{1,0}$ reversed too late for all of the flow profiles and parameters tried. And in simulations of Cycle 23 driven by idealised BMRs, \cite{Whitbread2017} found that a decay term with $\tau < 5\,\mathrm{yr}$ helped to reduce unrealistically high values of $b_{1,0}$. However, they found that emerging active regions with observed shapes reduced $b_{1,0}$ in itself \citep[as did][for Cycle 24]{Yeates2020}, and the optimization did not strongly select for a particular $\tau$. Moreover, the fit of the optimum model did not improve significantly when the decay term was included in the model compared to when it was not. \cite{Lemerle2015} also found that $\tau$ was not strongly constrained by the optimization process, with acceptable solutions found for suitable parameter combinations with $\tau$ in the range from 7--32$\,\mathrm{yr}$. In summary, the presence of a decay term as required by \cite{Schrijver2002} does not seem to be ruled out by observations.

It should be noted that, in principle, an additional decay term is not the only way to reduce the cycle-to-cycle memory of $b_{1,0}$ in the model. Alternatives that have been adopted include imposed cycle-to-cycle variations in either the meridional flow speed \citep{Wang2002} or the tilt angles of emerging BMRs \citep{Cameron2010}. It is difficult to choose definitively between these options with only about four solar cycles of full magnetogram observations.


\subsection{Diffusive Interior Model} \label{sec:baumann}

An improved parametrization for \eqref{eqn:rdiff} was suggested by \cite{Baumann2006}. They observed that if one assumes a purely diffusive evolution with uniform diffusivity $\eta=\eta_0$ throughout the convection zone, then the term $R_\eta$ may be approximated using only $B_r$ on the solar surface.

Specifically, \cite{Baumann2006} consider a purely poloidal field ${\bf B} = \nabla\times\nabla\times\big({\bf r}P\big)$ inside the convection zone $R_{\rm b}<r<R_\odot$, with boundary conditions $B_r(R_{\rm b},\theta,\phi)=0$ and $B_\theta(R_\odot,\theta,\phi)=B_\phi(R_\odot,\theta,\phi)=0$. Under a purely diffusive decay
\begin{equation}
    \frac{\partial{\bf B}}{\partial t} = -\eta_0\nabla\times\big(\nabla\times{\bf B}\big)
    \label{eqn:diffcz}
\end{equation}
with $\eta_0$ constant, and a suitable \edit{gauge choice for $P$}, this reduces to the scalar problem
\begin{equation}
    \frac{\partial P}{\partial t} = \eta_0\nabla^2 P, \quad \left.\frac{\partial}{\partial r}\big(rP\big)\right\rvert_{r=R_\odot} \edit{=} \quad P\Big\rvert_{r=R_{\rm b}} = 0.
    \label{eqn:sdiff}
\end{equation}
The solution, omitting the monopole term, may be written as an expansion
\begin{equation}
    P(r,\theta,\phi,t) = \sum_{n=0}^\infty\sum_{l=1}^{\infty}\sum_{m=-l}^l \big[ a_{l,n}j_l(k_{l,n}r) + c_{l,n}y_l(k_{l,n}r)\big]Y_{l}^m(\theta,\phi)\mathrm{e}^{-\eta_0 k_{l,n}^2t},
    \label{eqn:sbau}
\end{equation}
where $Y_{l}^m$ are spherical harmonics and $j_l$, $y_l$ are spherical Bessel functions of the first and second kinds. Linearity of \eqref{eqn:sdiff} allows \edit{\cite{Baumann2006}} to set $a_{l,n}=1$ without loss of generality, so the inner boundary condition fixes the other coefficient
\begin{equation}
    c_{l,n} = -\frac{j_l(k_{l,n}R_{\rm b})}{y_l(k_{l,n}R_{\rm b})}.
\end{equation}
The upper boundary condition then gives
\begin{align}
    &l\big[j_l(k_{l,n}R_\odot)y_l(k_{l,n}R_{\rm b}) - y_l(k_{l,n}R_\odot)j_l(k_{l,n}R_{\rm b})\big] = \nonumber\\ &\qquad\qquad k_{l,n}R_\odot\big[j_{l-1}(k_{l,n}R_\odot)y_l(k_{l,n}R_{\rm b}) - y_{l-1}(k_{l,n}R_\odot)j_l(k_{l,n}R_{\rm b})\big].
\end{align}
This equation must be solved numerically for each $l$ and $n$ to determine the eigenvalues $k_{l,n}$, which give the decay times $\tau_{l,n} = (\eta_0 k_{l,n}^2)^{-1}$ for each component, where $l$ is the spherical harmonic degree and $n$ is the radial mode number. Since the SFT model does not give the subsurface radial structure, \cite{Baumann2006} propose to keep only the modes with $n=0$, which are the slowest decaying modes for each $l$. They modify the SFT equation \eqref{eqn:sft} to
\begin{equation}
        \frac{\partial B_r}{\partial t} + \nabla_h\cdot\big({\bf u}_h B_r\big) = \eta\nabla_h^2 B_r - \sum_{l=1}^\infty\sum_{m=-l}^l\frac{b_{l,m}(t)}{\tau_{l,0}}Y_l^m(\theta,\phi) + S,
        \label{eqn:sftbau}
\end{equation}
where $b_{l,m}(t)$ are the spherical harmonic coefficients in the expansion of $B_r$,
\begin{equation}
    B_r(\theta,\phi,t) = \sum_{l=1}^\infty\sum_{m=-l}^lb_{l,m}(t)Y_l^m(\theta,\phi).
\end{equation}
The interior diffusivity $\eta_0$ that determines $\tau_{l,0}$ is taken to be different from the coefficent $\eta$ of the classical diffusion term.

Note that, since radial modes with $n>0$ are neglected, the effect on $b_{1,0}$ is identical to the simple exponential decay term, with $\tau = \tau_{1,0} = (\eta_0 k_{1,0}^2)^{-1}$. Accordingly, \cite{Baumann2006} showed that their alternative form of the decay term can also reduce the spurious long-term memory of the SFT model, as illustrated in the middle and bottom rows of Figure \ref{fig:decay}. They found that diffusivity values in the range $\eta_0=50-100\,\mathrm{km}^2\mathrm{s}^{-1}$ gave polar field evolutions consistent with recent observations. For $R_{\rm b}=0.7R_\odot$, and since $k_{1,0}\approx 5.46$, this corresponds to decay times for $b_{1,0}$ in the range $\tau_{1,0}\approx 5-10\,\mathrm{yr}$.  In their model driven by idealized BMRs, \cite{Whitbread2017} found an optimum $\eta_0=190\,\mathrm{km}^2\mathrm{s}^{-1}$, giving a decay time $\tau_{1,0}=2.7\,\mathrm{yr}$, in agreement with the $\tau$ found by optimizing the simple exponential decay term. \cite{Virtanen2017} also adopted the \cite{Baumann2006} model, but in a simulation where active regions had observed shapes; they found a value $\eta_0=100\,\mathrm{km}^2\mathrm{s}^{-1}$ to give reasonable results.

It is worth remarking that these implementations of \eqref{eqn:sftbau} have used different diffusivities for $\eta$ (the classical horizontal diffusion) and $\eta_0$ (which determines $\tau_{l,0}$). Moreover, the extra term in \eqref{eqn:sftbau} includes both radial and horizontal diffusion due to the interior diffusivity $\eta_0$. If one evaluates the radial diffusion term \eqref{eqn:rdiff} for a single mode of the interior solution \eqref{eqn:sbau}, one obtains
\begin{equation}
    \frac{\eta_0}{R_\odot^2}\frac{\partial^2}{\partial r^2}\big(r^2B_r\big) = -\eta_0\left(k_{l,n}^2 - \frac{l(l+1)}{R_\odot^2}\right)B_r,
\end{equation}
giving a decay time $\tau_{l,n}'= \eta_0^{-1}[k_{l,n}^2 - l(l+1)/R_\odot^2]^{-1}$ for radial diffusion alone. However, for small $l$ the difference from $\tau_{l,n}$ is negligible.

\subsection{Other Turbulent Transport Effects} \label{sec:pumping}

If we drop the simple assumption of a turbulent diffusion in the mean-field induction equation \eqref{eqn:induc}, then there are a wealth of possible transport effects that could be explored in SFT models. One such effect -- expected to be present from numerical convection simulations -- is turbulent pumping \citep{Petrovay1994}, which adds $-\boldsymbol{\gamma}\times{\bf B}$ to the turbulent electromotive force (mathematically equivalent to ${\bf u}$). Downward pumping ($\gamma_r<0$) in a region near the surface could reduce the aforementioned diffusive link of active regions to deeper layers \citep{Cameron2012b, Karak2016}. This is because it will tend to make the magnetic field lines radial, and \edit{-- as noted earlier --} if $B_\theta,B_\phi\approx 0$ in some region near the surface then it follows from \eqref{eqn:extrar} that $R_\eta=0$, so no additional radial diffusion term should be included in the SFT model. Latitudinal pumping ($\gamma_\theta\neq 0$) is also found to be very strong in convection simulations. However, this relies on a significant influence of rotation on the turbulence, which is weaker nearer the surface than in deeper layers.

\section{Beyond the Classical Model}\label{sec:beyond}

Several  have sought to improve on the classical SFT model described in the previous sections. We therefore conclude this review by outlining some of these developments.

\subsection{Improved Small-Scale Flows}

The approximation of small-scale flows by a uniform supergranular diffusivity, $D$, is perhaps the greatest simplification in the classical model. Three main approaches for improving the fidelity of the small-scale flow model have been applied.

\begin{figure}
    \centering
    \includegraphics[width=\textwidth]{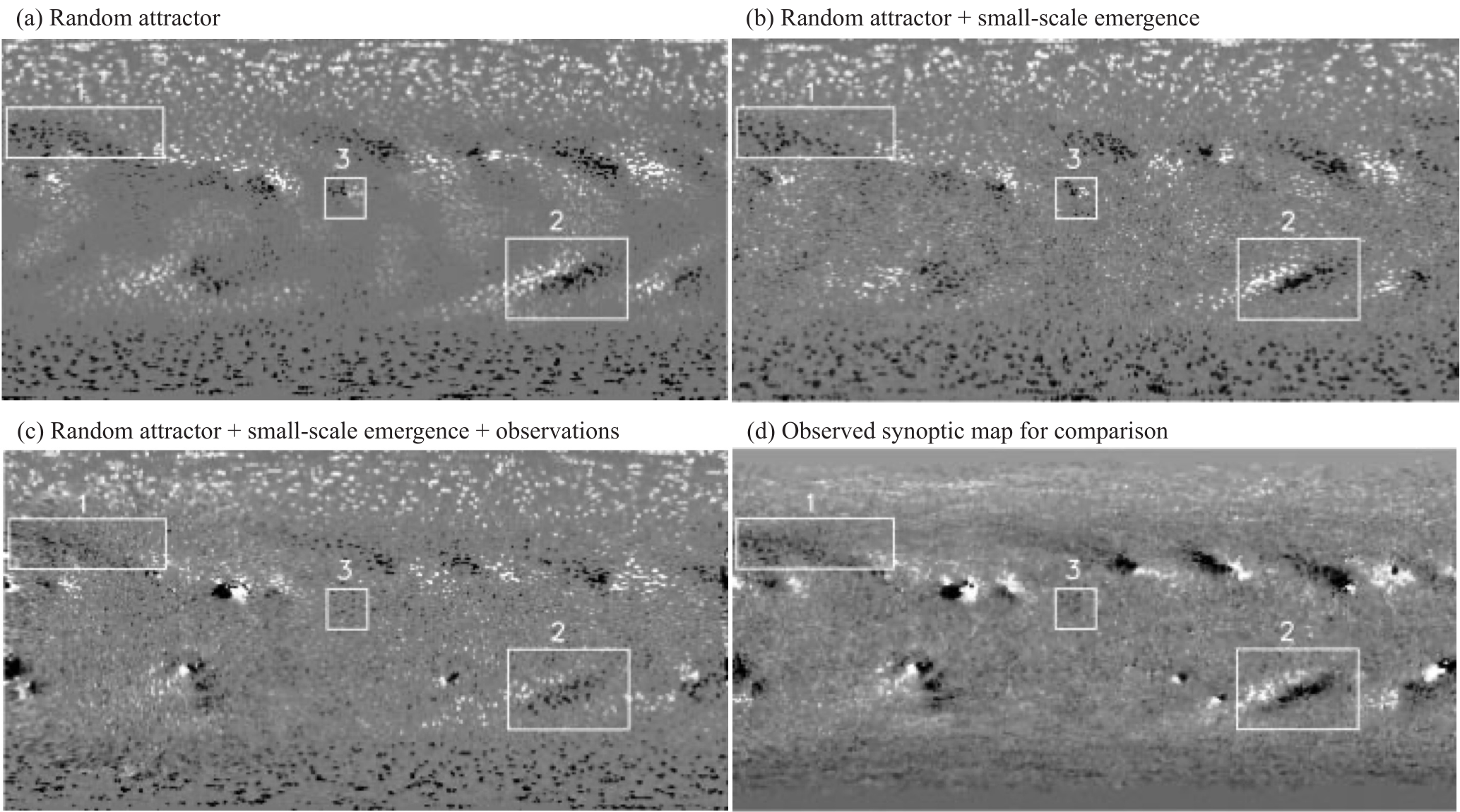}
    \caption{Illustration of the ``random attractor'' model for flux dispersal, taken from several figures of  \cite{Worden2000}. Panels (a)-(c) show simulated maps after evolving for 27 days, all starting from a synoptic map for CR1928 but with successively more model components included. (Differential rotation and meridional flow were included in all three cases.) Panel (d) shows the ``ground truth'': an observed synoptic map for CR1929. In all cases $B_r$ is shown in greyscale (white positive, black negative). (Reproduced with permission from Springer Nature. Original article: \url{https://doi.org/10.1023/A:1005272502885})}
    \label{fig:worden}
\end{figure}

Computationally cheapest is the method of \cite{Worden2000}, whose primary aim was to improve the unobserved or poorly observed regions of synoptic maps. For this application, the classical diffusion model is not ideal because it does not reproduce the ``clumping'' of magnetic flux on supergranular network boundaries that is clearly evident in observed portions of the map. To better reproduce this, \cite{Worden2000} replaced the diffusion with a ``random attractor'' term added to each pixel in the map (without increasing the resolution compared to the classical SFT model). This is shown in Figure \ref{fig:worden}(a). They also added a random emergence term to each pixel to sustain the small-scale background field. This background field was found not to affect the diffusion of large-scale flux patterns, but it gives a more accurate net flux in quiet regions (Figure \ref{fig:worden}b). The technique was successful in improving the appearance of simulated maps, and continues to be used in the Air Force Data-Assimilative Photospheric flux Transport model \citep[ADAPT;][]{Arge2010, Hickmann2015}.

A second approach is to dispense completely with parametrization of the small-scale flows, and model them directly through the advection term. This requires higher spatial and temporal resolution so as to resolve individual convective cells on the computational grid. Nevertheless, it has been applied successfully in the Advective Flux Transport (AFT) model \citep{Upton2014, Upton2018}. In this model, the small-scale flows are randomly imposed, based on a vector spherical harmonic decomposition of the form
\begin{align}
    u_\theta(\theta,\phi) &= \sum_{l=1}^{l_{\rm max}}\sum_{m=0}^l\left(S_l^m\frac{\partial Y_l^m(\theta,\phi)}{\partial\theta} + T_l^m\frac{1}{\sin\theta}\frac{\partial Y_l^m(\theta,\phi)}{\partial\phi}\right),\\
    u_\phi(\theta,\phi) &= \sum_{l=1}^{l_{\rm max}}\sum_{m=0}^l\left(S_l^m\frac{1}{\sin\theta}\frac{\partial Y_l^m(\theta,\phi)}{\partial\phi} - T_l^m\frac{\partial Y_l^m(\theta,\phi)}{\partial\theta}\right),
\end{align}
where the complex amplitudes $S_l^m$ and $T_l^m$ determine the curl-free and divergence-free components of ${\bf u}_h$ and are chosen to match the spectrum to observations. \cite{Hathaway2000} found that observed Doppler flows could be well matched by a two-component spectrum, comprising a supergranular component centred on $l=110$ and a granular component centered on $l=4000$.

The third approach is to dispense with a computational grid altogether and model the magnetic flux by a discrete ensemble of individual flux ``concentrations''. This was implemented by \cite{Schrijver2001} whose main aim was to simulate cool stars other than the Sun, and who therefore wanted to include the mixed-polarity network of small-scale magnetic flux because of its contribution to chromospheric emission. The discrete model of \cite{Schrijver2001} includes (i) emergence of both active regions and ephemeral regions as BMRs, (ii) a large-scale random walk dispersal as well as differential rotation and meridional flow, (iii) a model for fragmentation and coalescence of flux concentrations, and (iv) cancellation of flux between opposite polarity fragments. The model has been successfully applied over all latitudes \citep{Schrijver2001b} and over a full 11-year cycle \citep{Schrijver2008}. A similar model in Cartesian geometry was applied by \cite{MartinBelda2016} to study the dispersion of a single active region.

One notable new feature that all three of these models have in common is nonlinearity: the rate of magnetic flux dispersal is chosen to depend on the local magnetic field strength, $\lvert B_r\rvert$. In particular, dispersal is suppressed in strong-field regions, compared to the classical diffusion model. This better represents real active regions which suppress shedding of the magnetic flux by supergranulation \citep{Schrijver1989}. The effect is particularly important for more active stars \citep{Schrijver2001} but is still clearly observed on the Sun.

\subsection{Fluctuating Large-Scale Flows}

The classical model neglects fluctuations in the meridional flow and differential rotation, keeping them steady for periods of a solar cycle or longer. However, observations do suggest variations over the course of the cycle, particularly in the meridional flow. For example, \cite{Hathaway2010} estimated the flow from cross-correlating latitudinal strips in magnetograms over Solar Cycle 23, and found that the dominant Legendre component, $P_2^1\sim\sin(2\theta)$, reduced in amplitude from $11.5-13\,\mathrm{m}\mathrm{s}^{-1}$ at cycle minimum to only $8.5\,\mathrm{m}\mathrm{s}^{-1}$ at cycle maximum.

A plausible cause of meridional flow variations is the observed inflow toward active regions determined by helioseismology \citep{Gizon2001}. In SFT simulations, \cite{Jiang2010b} showed that an axisymmetric meridional inflow toward the activity belts leads to a significant decrease of the polar field, suggesting that such meridional flow variations could be a significant ingredient in the SFT model. And \cite{Cameron2010} pointed out that the variations in $P_2^1$ found by \cite{Hathaway2010} could be explained by this inflow, without the need for an overall modulation of meridional flow speed.

Other studies have accounted for the observed dependence of inflow speed on the active region magnetic flux, through applying a nonlinear velocity that depends on $\lvert B_r\rvert$. \edit{Whilst more detailed models for magnetic back-reaction on flows and transport coefficients have been introduced in dynamo models \citep{Rempel2006}, SFT studies have so far been limited to simple parametrizations.} 
\cite{Derosa2006} added a velocity of the form
\begin{equation}
    \delta{\bf u}(\theta,\phi,t) = \alpha\nabla\lvert\overline{B_r}\rvert^\beta 
\end{equation}
to the discrete SFT model -- where $\overline{B_r}$ denotes a Gaussian smoothing of the original $B_r$ with width $15^\circ$ -- but found that the observed flow speeds ($~50\,\mathrm{m}\mathrm{s}^{-1}$) prevented altogether the dispersal of active regions. However, \cite{MartinBelda2016} did not find this problem and proposed that the original calculations of \cite{Derosa2006} were underestimating the flux dispersal because they continued to apply the nonlinear damping of dispersal within the active region, while the inflows alone could themselves account for the damping effect. \cite{Cameron2012} proposed an axisymmetric parametrization
\begin{equation}
\delta u_\theta(\theta, t)=c_0\int_0^\pi\frac{\sin(\theta')}{\sin(30^\circ)} \frac{\mathrm{d}\langle\lvert B_r\rvert\rangle}{\mathrm{d}\theta'}\mathrm{e}^{-(\theta-\theta')^2/\sigma}\,\mathrm{d}\theta',
\end{equation}
which corresponds to a Gaussian smoothing of the derivative in latitude (with $\sigma$ chosen to give width $20^\circ$). The $\sin(\theta')$ factor suppresses unrealistically strong fluctuations at high latitudes, and an amplitude $c_0=9.2\,\mathrm{m}\,\mathrm{s}^{-1}\mathrm{G}^{-1}$ gives comparable inflow speeds to \cite{Gizon2001}.
Again, the presence of inflows reduces the axial dipole at the end of the solar cycle, by about $30\%$ in a moderate cycle \citep{MartinBelda2017}, with about a $9\%$ variation between cycles suggesting that this nonlinearity could conceivably help to saturate the Babcock-Leighton dynamo.
\citet{Nagy2020} coupled an SFT model with flux-dependent inflows to such a dynamo model. They confirmed that inflows do indeed tend to have a stabilizing effect on cycle amplitudes, although they also greatly increase the probability of the dynamo entering a grand minimum of reduced activity -- a nonlinear effect which is not apparent from SFT alone.
On the other hand, \cite{Yeates2014} found that the inflows in a BMR-driven SFT model for Cycle 23 gave poorer matches to the observed butterfly diagram and dipole reversal time.

A more pragmatic approach is to impose the observed flow variations directly, as in the AFT model \citep{Upton2014}, where the best-fit Legendre coefficients are extracted from 27-day averaged velocity fields derived from magnetogram cross-correlation. These then determine ${\bf u}_h(\theta,t)$ in the model, allowing variations in both meridional flow and differential rotation. \edit{Using data from Solar Cycle 23, \cite{Upton2014b} found that the fluctuating meridional flow in the AFT model actually increased the axial dipole strength by $20\%$ compared to a simulation where the meridional flow was fixed to a steady latitudinal profile. Thus it is possible that meridional flow variations can increase the axial dipole as well as reduce it.}

\subsection{Observational Data Assimilation}

In applications where the aim is to recreate as accurately as possible the real Sun at an observed time, it makes sense to construct magnetic maps that combine SFT model results with real observations. The role of the SFT model is then to fill in unobserved (or poorly observed) parts of the solar surface, such as high latitudes or the far side of the Sun.
This approach is central to the model of \cite{Worden2000}, as illustrated in Figure \ref{fig:worden}(c) which shows the result of combining daily magnetogram observations with the simulation. The observations are weighted more highly near disk-centre and also eastward of Central Meridian (where the time since previous observation is greatest). Similar assimilation of observed magnetograms has been applied in the discrete SFT model \citep{Schrijver2003} and in the AFT model of \cite{Upton2014}.

A more sophisticated approach to data assimilation has been implemented in the ADAPT model, which includes several different sequential data-assimilation methods such as ensemble Kalman filtering \citep{Hickmann2015}. The concept is to perform an ensemble of model runs. Each is adjusted at intervals using the observed magnetogram data, with observations being given greater weight in areas where the model runs disagree with one another.

Unfortunately, difficulties arise in driving time-dependent coronal magnetic field simulations from SFT models with data assimilation. In such simulations, the required photospheric boundary condition is the tangential electric field ${\bf E}_h$, not simply $B_r$. In the classical SFT model, the natural electric field would be
\begin{equation}
    {\bf E}_h = -{\bf u}\times{\bf B} + \eta\nabla\times{\bf B} + {\bf E}_S,
\end{equation}
where ${\bf E}_S$ accounts for the source term (\textit{i.e.}, $-{\bf e}_r\cdot\nabla\times{\bf E}_S = S$).
When $S$ comprises individual active regions that have no net magnetic flux, a well-behaved electric field can be determined \citep[\textit{e.g.},][]{Yeates2022}. But if the magnetic flux is unbalanced over a larger region then it is impossible to find a localized ${\bf E}_S$ as would be expected from Ohm's Law \citep{Yeates2017}. This can be a problem when observed magnetograms are incorporated directly, particularly when active regions straddle the edge of the assimilation region so that only one polarity is included. If the flux imbalance is corrected by spreading it over the full Sun, the resulting spurious electric fields lead to generation of significant spurious electric currents in time-dependent coronal simulations \citep{Weinzierl2016}. \edit{Of course, this problem is not restricted to data assimilation, but could arise from the use of any unbalanced source term.}

In practice the simplest way to ensure flux balance is to rephrase the right-hand side of equation \eqref{eqn:sft2d} as $-{\bf e}_r\cdot\nabla\times{\bf E}_h$, then apply a ``constrained transport'' discretization with a staggered mesh \citep{Yee1966}. Here $E_\theta$ and $E_\phi$ are defined at cell edges, and $B_r$ at cell centres. Such a numerical scheme is used, for example, by \cite{Yeates2014}. When assimilating magnetograms into the SFT model in this framework, one would estimate ${\bf E}_h$ from the observed front-side evolution. In the case of a flux imbalance, this would automatically create a balancing polarity just outside the observed region, minimizing disruption to the global topology of the coronal magnetic field. However, it remains the case that systematic errors in observed magnetograms, especially centre-to-limb variations of the errors, are not well understood. A better understanding of these errors will require forward modelling with radiative MHD and Stokes polarimetric inversions.

\edit{A final remark is that the simplified decay term $B_r/\tau$ from Section \ref{sec:exp} may also be written as the curl of an electric field. In particular, we would need ${\bf e}_r\cdot\nabla\times{\bf E}=B_r/\tau$. For example, writing ${\bf E}=-\nabla\times(\Psi{\bf e}_r)$, we could determine $\Psi$ and hence ${\bf E}$ by solving the Poisson equation
\begin{equation}
    \nabla_h^2\Psi = \frac{B_r}{\tau},
\end{equation}
which has a unique solution on the sphere since $\int_SB_r\,\mathrm{d}S=0$. Of course, this does not mean that this approximation is a good representation of the real radial diffusion term \eqref{eqn:rdiff}; for example, this particular ${\bf E}$ \edit{will not be} localized to the active region itself.}

\backmatter

\bmhead{Supplementary information}

None.

\bmhead{Acknowledgments}

We thank the International Space Science Institute for supporting the workshop where this review originated. ARY was supported by STFC (UK) consortium grant ST/W00108X/1. JJ was supported by the National Natural Science Foundation of China (grant Nos. 12173005 and 11873023). KP acknowledges support by the European Union's Horizon 2020 research and innovation programme under grant agreement No.~955620.
The collaboration of the authors was also facilitated by support from the
International Space Science Institute through ISSI Team 474.
The SDO data used in Figures \ref{fig:paper_opt_bfly}, \ref{fig:paper_noflow} and \ref{fig:non-bmr} are courtesy of NASA and the SDO/HMI science team. \edit{We thank the two anonymous reviewers for improving the article.}

\section*{Declarations}

\textbf{Competing interests.} The authors have no competing interests to declare that are relevant to the content of this article.

\bibliography{sft-bibliography}

\end{document}